\documentclass[prx,aps,twocolumn,superscriptaddress,amsmath,amssymb,floatfix,citeautoscript]{revtex4-2}

\usepackage{etoolbox}

\makeatletter
\patchcmd{\subsubsection}{\itshape}{\bfseries\itshape}{}{}
\makeatother

\usepackage{graphicx}
\usepackage{dcolumn}
\usepackage{bm}

\usepackage{color}
\usepackage{bbm}
\usepackage{amsfonts,amsmath,amssymb,stmaryrd}
\usepackage{bbold}    
\usepackage{todonotes} 
\usepackage[normalem]{ulem} 
\usepackage[nolist]{acronym}
\usepackage{orcidlink} 

\newcommand{\tr}{\text{Tr}}

\newcommand{\rh}{\hat{\rho}}

\newcommand{\id}{\hat{\mathbbm{1}}}

\newcommand{\half}{\frac{1}{2}}
\renewcommand{\vec}[1]{\bm{#1}}

\newcommand{\D}{\hat{\Delta}}

\newcommand{\Y}{\text{Y}}
\newcommand{\td}{\text{d}}

\newcommand{\Sc}{{\mathcal{S}}}

\newcommand{\Om}{\mathbf{\Omega}}
\newcommand{\Omt}{\tilde{\mathbf{\Omega}}}
\newcommand{\Dm}{\hat{\mathbf{\Delta}}}

\newcommand{\W}{{\cal W}}

\graphicspath{{figures/}}

\newcommand{\mfl}[1]{\textcolor{blue}{#1}}

\begin{document}


\title{Truncated Wigner Approximation for Spins in Continuous Phase Space
}

\author{Jens Hartmann \orcidlink{0009-0002-9563-2491}}
\email{jhartman@rptu.de}
\author{Tom Schlegel \orcidlink{0009-0003-0204-5712}}
\affiliation{Department of Physics and Research Center OPTIMAS, RPTU University Kaiserslautern-Landau, D-67663 Kaiserslautern, Germany}
\author{Viktoria Noel \orcidlink{0009-0006-5361-653X}}
\affiliation{Institut f\"ur Theoretische Physik and Center for Integrated Quantum Science and Technology (IQST),
Eberhard Karls Universit\"at T\"ubingen, D-72076, T\"ubingen, Germany}
\author{Christopher D. Mink \orcidlink{0000-0003-1867-4313}}
\author{Michael Fleischhauer \orcidlink{0000-0003-4059-7289}}
\affiliation{Department of Physics and Research Center OPTIMAS, RPTU University Kaiserslautern-Landau, D-67663 Kaiserslautern, Germany}

\date{\today}

\begin{abstract}
We review the truncated Wigner approximation (TWA) for spins as a computationally inexpensive numerical approximation method to describe interacting and / or dissipative many-body spin systems.  Using the Wigner-Moyal mapping from Hilbert space to a suitable phase space, the many-body density matrix is represented by a c-number distribution, the Wigner function. The gauge freedom in continuous phase space can be exploited to find positive Wigner functions for a large class of spin states, including entangled ones.
Employing different sets of correspondence rules,
we derive equations of motion for the Wigner function,
which, applying controlled approximations, can be mapped to stochastic differential equations. This allows a computationally inexpensive simulation of expectation values. Using a phase-space analog of the quantum regression theorem also multi-time correlations and spectra can be obtained.
To illustrate the potential of the method, we benchmark the TWA for spins with some exactly solvable problems of interacting, dissipative spin systems, and then discuss its application to collective processes, such as the superradiant emission of  light.
Extending the TWA to imaginary time furthermore provides a tool to approximately calculate thermal and ground states of spin Hamiltonians. 
Finally, we show that the TWA stochastic equations can equivalently be derived within a path-integral approach, provided that the operator products in the dissipator are rigorously mapped onto the curved phase space.
\end{abstract}
\maketitle

\begin{acronym}

\acro{twa}[TWA]{truncated Wigner approximation}

\acro{dtwa}[DTWA]{discrete truncated Wigner approximation}

\acro{itwa}[iTWA]{imaginary-time truncated Wigner approximation}

\acro{qmc}[QMC]{Quantum Monte Carlo}

\acro{mps}[MPS]{matrix-product state}

\acro{pde}[PDE]{partial differential equation}

\acro{fpe}[FPE]{Fokker-Planck equation}

\acro{sde}[SDE]{stochastic differential equation}

\acro{obe}[OBE]{optical Bloch equation}

\acro{eom}[EOM]{equation of motion}

\acro{ode}[ODE]{ordinary differential equation}

\end{acronym}

\section{Introduction}
\begin{figure}[h]
    \centering
        \includegraphics[width=0.48\textwidth]{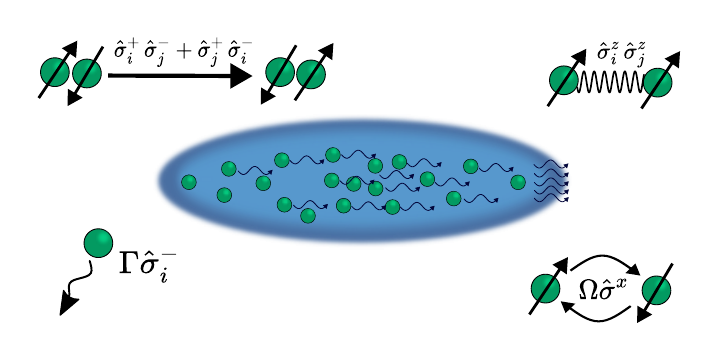}
        \vspace{-5mm}
        \caption{The microscopic basis of 
        a large class of quantum-many-body problems, ranging from magnetism to collective light emission, are interacting, open spin systems. The faithful description of their dynamics in large-scale systems remains, however, a major challenge.}
        \label{fig:system}
\end{figure}

The accurate description of the many-body dynamics of interacting, open spin systems (see Fig.~\ref{fig:system}) remains one of the major challenges of theoretical physics. Exact numerical treatments are limited to small systems or to the classical limit where the dynamics can be fully described by excitation probabilities and large systems can be tackled using classical Monte Carlo simulations~\cite{binder2005monte,voter2007radiation}. 
\ac{qmc} techniques~\cite{suzuki1976relationship,suzuki1977monte,foulkes2001quantum}, on the other hand,  can only be applied efficiently if the system is frustration free~\cite{Troyer-PRL-2005}. 
Therefore, there is an active and ongoing quest for approximate, beyond mean-field calculation methods. Extensions such as cluster-mean-field~\cite{jin2016cluster}, variational approaches~\cite{weimer2015variational}, or cumulant expansions~\cite{kubo1962generalized,kramer2015generalized,robicheaux2021beyond,plankensteiner2022quantumcumulants,rubies2023characterizing,masson2024dicke,Kerber2025} often give good results for certain problems but are in general not well controlled and 
numerically expensive, unless special symmetries can be exploited~\cite{holzinger2025symmetry}.
Methods based on \ac{mps} expansions of the density matrix~\cite{vidal2004efficient,verstraete2004matrix,verstraete2008matrix,schollwock} or variational \ac{mps} techniques~\cite{cui2015variational} are generally limited to one spatial dimension with short-range couplings and small evolution times. 

In the present paper, we give a detailed review of an approximated simulation method, the \ac{twa} for spins.  It is a phase-space technique that exploits mapping of the Hilbert space of interacting, dissipative spins to a continuous phase space, based on the Stratonovich-Weyl correspondence~\cite{Brif-PRA-1999}. 
Phase-space approaches
recast the quantum dynamics of the many-body density matrix as a \ac{pde} of a quasiprobability distribution, which represents the quantum state.
Approximating this \ac{pde} by a \ac{fpe} allows to 
simulate the quantum dynamics in an efficient way by \acp{sde}. Expectation values of operators can then simply be obtained as statistical averages.
Phase-space representations of Hilbert spaces are not unique, as relating operator products to expressions in phase space requires fixing a specific operator ordering, and different approaches can strongly vary in the quality of the approximation results.
Clearly, an exact mapping of the quantum many-body dynamics to a
classical, nonlinear diffusion process in phase space is, in general, not possible. However, selecting suitable representations allows one to capture the most important quantum effects. The challenge is to find a representation, for which stochastic simulations have the highest accuracy.
In this search, one has to take into account that a faithful description of quantum dynamics by stochastic evolution is only possible if the initial and final states have (nearly) positive definite quasi-probability distributions. 

In quantum optics phase-space approaches have successfully been employed to study open, interacting bosonic systems, such as lasers, nonlinear optical processes, or ultra-cold quantum gases~\cite{gardiner2004quantum}. 
Although descriptions of quantum spins in phase space have been well known since the late 1950s from the early works of Stratonovich~\cite{stratonovich1957distributions}, see e.g.~\cite{varilly1989moyal,Brif-PRA-1999}
for reviews, only more recently phase-space representations have been used to describe the time evolution of interacting spin systems.
Here, two different descriptions have to be distinguished: (i) representations in a discrete space
as introduced by Wootters~\cite{Wootters,DTWA} and (ii) continuous SU(2) representations
\cite{stratonovich1957distributions,varilly1989moyal,Brif-PRA-1999}.

In~\cite{DTWA}
a semi-classical approach in discrete phase space -- the \ac{dtwa}  -- was introduced.
Here, the many-body dynamics is described by a set of classical equations of motion of discrete phase points at the mean-field level. Similarly to the \ac{twa} for Bose fields~\cite{Steel,gardiner2004quantum,Blakie-AdvPhys-2008,Polkovnikov-AnnPhys-2010}, quantum fluctuations are partially incorporated by sampling an initial discrete distribution~\cite{Wootters}.
The \ac{dtwa} can predict collective observables of unitary interacting spin-$\frac{1}{2}$ systems rather accurately. However, its applicability is often limited to short or intermediate time scales~\cite{czischek2019quenches,PhysRevA.99.043627,khasseh2020discrete,kunimi2021dtwaBenchmark,zhu-NJP-2019}. 
In Ref.~\cite{huber2021realistic} dephasing was incorporated by Markovian classical noise fields coupled to the $x,y$ components of the spins, turning the deterministic equations of motion into stochastic ones.
In contrast, decay and incoherent drive are associated with non-classical noise in standard \ac{dtwa}, which prevents numerical simulations using \acp{sde} and therefore further approximations were made~\cite{huber2021realistic,Huber-SciPost-2021}. 
In Ref.~\cite{Singh2021} an open-system version of the \ac{dtwa} was proposed by introducing the spin norm as an ad-hoc additional degree of freedom. 
For a single spin, very good agreement with exact results was found, but the 
range of validity of the approach and its applicability to more general reservoir couplings remains questionable.

A stochastic description of open, interacting spin-$1/2$ systems in continuous phase space, termed continuous \ac{twa} or short \ac{twa} for spins, was first suggested in~\cite{MinkPRR2022}. The continuous SU(2) representation of the quantum state has the advantage of providing a straightforward way to derive equations of motions for the Wigner quasi-probability distribution with the help of a set of simple 
correspondence rules, which are 
operator-differential mappings. The same rules allow to directly extend the approach to incorporate dephasing and decay. These mappings yield an exact \ac{eom} for the continuous SU(2) Wigner function. Truncating the PDE to second order with positive definite diffusion then yields a Fokker-Planck equation (FPE), which is equivalent to a set of Ito stochastic differential equations. These can be simulated efficiently. 
In contrast to the discrete approaches, mentioned above, the single-particle dynamics of spins including coherent driving, e.g. by magnetic fields, as well as decay, dephasing, and incoherent pump processes are described exactly.
Most importantly, the continuous phase-space representation has a gauge degree of freedom~\cite{MinkPRR2022}.
This makes it possible to describe a much larger class of spin states in terms of a positive quasiprobability distribution.
As a consequence, the continuous \ac{twa} for spins can accurately describe some processes where the discrete quasi-probability distribution of the final state would not be positive definite (see Sec.~\ref{sec:SDEs} C). 

Moreover, correspondence rules between Hilbert and phase space are not unique either and can be adjusted to the specific problem.
A set of approximate correspondence rules different from those used in~\cite{MinkPRR2022} was derived in~\cite{mink2023collective}, which is better suited for collective spin phenomena, such as superradiant emission of light.
The collective correspondence rules have since been successfully applied to describe experimental observations of collective light emissions of extended atomic ensembles in chiral~\cite{tebbenjohannsPRA2024} and non-chiral~\cite{spahn2026motioninduceddirectionalitycollectiveemission} waveguides, but do not describe sub-radiant states.

In the following, we will present a detailed overview of the \ac{twa} for spins, introduce some important formal extensions, and provide specific applications, highlighting the potentials and limitations of the method.
The paper is organized as follows: For a reader that wants to obtain a fast overview of the method without derivations and background, we will give an ''executive summary'' of the method in Sec.~\ref{sec:overview}. We will then introduce the continuous SU(2) phase-space representation of spins in Sec.~\ref{sec:Phase-space} in detail, explain its gauge freedom and demonstrate how the latter can be exploited to obtain positive quasiprobability distributions even of entangled states such as Bell states. In Sec.~\ref{sec:FPE} we will derive the EOM for the Wigner distribution for unitary and dissipative couplings, introducing both single-spin and collective correspondence rules.
Truncating these at the Fokker-Planck level, we derive stochastic differential equations for different interaction problems and collective problems in Sec.~\ref{sec:SDEs}.
How to calculate two-time correlation functions by transferring the quantum regression theorem to phase space is discussed in Sect.~\ref{sec:Two-time}, including some benchmark examples.
In order to calculate finite-temperature or ground states of many-body Hamiltonians, it is useful to extend the \ac{twa} for spins to imaginary time. This will be done in Sec.~\ref{sec:iTWA}. For bosonic fields there is a one-to-one correspondence between phase-space approaches and Keldysh path integrals approximated to second order in quantum fluctuations~\cite{Polkovnikov-AnnPhys-2010,sieberer2016keldysh}, which provides an alternative way to derive stochastic equations.
Very recently, a similar approach was put forward for 
spins in~\cite{hosseinabadi2025user}, which, however, contains inconsistencies in the treatment of decay processes, which leads to some errors in the simulation of open systems. 
In Sec.~\ref{sec:Keldysh} we show that a consistent path-integral description can also be obtained for spins, provided the operator products in the dissipator are mapped correctly onto the curved phase space using the Wigner--Moyal mapping introduced in Sec.~\ref{sec:Phase-space}. This approach recovers the EOMs obtained in Sec.~\ref{sec:FPE}, and a detailed description of this will be given in a separate publication~\cite{Noel:2026jra}.

\section{TWA in a nutshell}\label{sec:overview}

As first shown by Wigner~\cite{wigner1932quantum} in his studies of quantum corrections at thermodynamic equilibrium, and later expanded in the seminal work of Moyal~\cite{moyal1949quantum}, quantum mechanics, typically formulated in terms of states and operators in Hilbert space, can be equivalently represented in a classical phase space. Here, observables are expressed in terms of classical c-number functions and quantum states by quasi-statistical representations.
For spin-$1/2$ systems, using a continuous SU(2) representation~\cite{Brif-PRA-1999,varilly1989moyal},
all operators including the density matrix are expanded in an over-complete set of basis operators, called \textit{phase-point operators}.
They define the degrees of freedom of every spin by angles $\theta$ and $\phi$ on an enlarged Bloch sphere of radius $\sqrt{3}$
    \begin{align}
    \hat{\Delta}(\theta, \phi) =\half
    \begin{pmatrix}
        1 + \sqrt{3} \cos \theta && \sqrt{3} e^{-i \phi} \sin \theta\\
        \sqrt{3} e^{i \phi} \sin \theta && 1 - \sqrt{3} \cos \theta
    \end{pmatrix}. \label{eq:originKernel}
    \end{align}
The enlarged radius is needed to account for uncertainty relations of the quantum spin components. The coefficients for observables, ${\cal A}$, are called Weyl symbols, the coefficient of the density matrix, ${\cal W}$, the Wigner function
    \begin{equation}
        \hat A = \int \!\! d\Omega\,\, {\cal A}(\Omega)\, \hat{\Delta}(\Omega),\qquad \hat \rho = \int \!\! d\Omega\,\, {\cal W}(\Omega)\, \hat{\Delta}(\Omega),
    \end{equation}
    where $\Omega=(\theta,\phi)$.
    Expectation values of operators are evaluated by averaging the corresponding Weyl symbol with the Wigner function
    \begin{equation}
    \langle \hat A (t)\rangle = \int \!\! d\Omega\,\,   {\cal A}(\Omega)\, {\cal W}(\Omega,t).
    \end{equation}
Since the phase-point operators contain only spherical harmonics $Y_{l}^m(\theta,\phi)$ with $l=0,1$, adding any terms proportional to higher spherical harmonics to
the Wigner function, will not affect physical observables. This constitutes a \textit{gauge freedom} in the choice of ${\cal W}(\Omega)$, which can be exploited to find strictly 
positive Wigner functions, allowing one to interpret it as a probability distribution, which can be sampled. 
    
The aim of a phase-space method is to map the \ac{eom} of the quantum state $\hat \rho$, which we assume here to be of the Lindblad master equation type, to stochastic evolution equations. In a first step, an EOM for the Wigner function is obtained by
\begin{eqnarray*}
    &\frac{d}{dt} \rh =  -i \bigl[H,\rh\bigr] + \sum_j \Bigl(\hat L_j \rh\hat L_j^\dagger -\frac{1}{2}\bigl\{\hat L^\dagger_j\hat L_j,\rh\bigr\}\Bigr) 
= {\cal L}\rh\nonumber &\\
&\Updownarrow & \\
 &\int\!\! d\Omega \, \frac{\partial}{\partial t}  \W(\Omega,t) \, \hat \Delta = \int\!\! d\Omega\,  \W(\Omega,t) \, {\cal L}\hat \Delta&.
\end{eqnarray*}
The action of the superoperator ${\cal L}$ on the right hand side 
corresponds to products of Pauli operators acting on $\hat \Delta$. Being just products of $2\times 2$ matrices these terms 
can be expressed in terms of $\hat \Delta$ and its derivatives up to second order, which form a linear independent set of $2\times 2 $ matrices
\begin{equation}
    {\cal L} \hat\Delta \, \, \longleftrightarrow \, \, \hat \Delta, \, \, \partial_\phi \hat\Delta , \, \partial_\theta \hat\Delta, \, \, \textrm{and}\, \,\partial^2_\phi \hat \Delta, \,\,  \textrm{or} \,\,  \partial^2_\theta \hat \Delta.\nonumber
\end{equation}
This gives rise to a set of \textit{direct correspondence rules} between operators and c-numbers~\cite{MinkPRR2022}, given in detail later. 

If the Lindblad-superoperator ${\cal L}$ contains collective coupling terms, it is advantageous to apply different, approximate \textit{collective correspondence rules}, as derived in
\cite{mink2023collective}, which will also be discussed in detail later.

In both cases, partial integration leads to a partial differential equation for the Wigner function. 
To avoid complications arising from the fact that the curved phase space integration measure $d\Omega = d\phi\, d\theta\, \sin\theta$ contains the term ''$\sin\theta$'', one here uses the \textit{flattened} Wigner function
$\chi(\Omega,t) = \W(\Omega,t) \pi \sin\theta$ instead of $\W(\Omega,t)$.
Neglecting all higher-order derivatives in the partial differential equation for $\chi(\Omega,t)$, and approximating the coefficient-matrix of the second-order terms by a positive one,  results in a Fokker-Planck equation, which is equivalent to a set of Ito stochastic differential equations for the spin angles $\theta(t)$ and $\phi(t)$.
Their initial values must be sampled from the Wigner function of the initial state $\W(\Omega,t=0)$.
By exploiting its gauge freedom, a positive representation can be chosen for most relevant states.
The resulting equations are of the form
\begin{eqnarray}
d \theta_j &=& A_{\theta_j} \, dt  + \sum_l \Bigl( B_{\theta_j\theta_l}\, d W_{\theta_l}   + B_{\theta_j,\phi_l} \, d W_{\phi_l}\Bigr),\label{eq:SDE-1}\\
d \phi_j &=& A_{\phi_j}\, dt  + \sum_l \Bigl(B_{\phi_j\theta_l}\, d W_{\theta_l}   + B_{\phi_j,\phi_l} \, d W_{\phi_l}\Bigr),\nonumber 
\end{eqnarray} 
where the drift terms, $A_x=A_x(\Omega_n)$, and the noise terms, $B_{xy} (\Omega_n)$, are functions of all $\theta_n$ and $\phi_n$, and 
$dW_x= dW_x(t)$ are independent normalized Wiener processes, obeying
\begin{equation}
\overline{ \, d W_x(t) \, } = 0,\qquad \overline{\, d W_x(t) d W_y(t^\prime) \, } = \delta_{x,y} \delta(t-t^\prime) \, dt
\end{equation}
The precise form of the drift and diffusion terms for a given Liouvillian depend on the chosen correspondence rules, and will be discussed in detail later.
Expectation values of a state $\hat \rho(t)$ at time $t$ can then be evaluated by simulating the coupled \acp{sde}, eq.~\eqref{eq:SDE-1}, with initial values sampled from the Wigner function $\W(t=0)$ of the initial state
\begin{equation}
  \langle \hat A (t)\rangle = \int \!\! d\Omega\,  {\cal A}(\Omega)\, \W(\Omega,t) = \overline{\overline{ \, {\cal A}(\Omega(t))\, }}.
\end{equation}
The double line indicates averaging over the noise realizations ($d W_x$) and sampling of the initial distribution ($\W(\Omega,t=0)$). In the following, we will use a single overline, however, for notational simplicity.
The total number of equations to be solved for $N$ spins is $2N$ and thus the numerical effort scales only linearly with the system size.
The numerical simulations can be performed in Julia e.g. using the \texttt{DifferentialEquations.jl} package. The \acp{sde} are evolved using the Euler-Maruyama method as well as higher-order \ac{sde} solvers depending on the problems noise type.
In order to obtain good approximations, in general, a larger number of trajectories $N_\textrm{traj}$ is needed, which may also scale with the system size. For many applications this scaling is at most linear or a low-power polynomial, in some cases it can however also be exponential \cite{schlegel2026imaginarytimeevolutioninteractingspin}, in which case the \ac{twa} simulation becomes algorithmically hard.

\section{Continuous phase space representation  of spins} \label{sec:Phase-space}


\subsection{Continuous operator basis and mapping from Hilbert to phase space}

\subsubsection{Phase-point operators and Weyl symbols}

The over-complete  basis set in the $SU(2)$ representation of single-spin operators, eq.~\eqref{eq:originKernel},
can be expressed in terms of Pauli matrices
\begin{align}
    \hat{\Delta}(\theta, \phi) = \half \left[\id_2 + \vec{s}(\theta, \phi) \hat{\vec{\sigma}}\right], 
  \end{align}
with the c-number vector
\begin{align}
    \vec{s}(\theta, \phi) = \sqrt{3} (\sin \theta \cos \phi, \sin \theta \sin \phi, \cos \theta)^T. \label{eq:CartesianSpin}
\end{align}
It is a representation of a point on the surface of the sphere with radius $\sqrt{3}$. 
For a system of $N$ spin 1/2 particles holds
\begin{equation}
    \hat{\mathbf{\Delta}}(\mathbf{\Omega}) = \prod_{j=1}^N \hat{\Delta}_j(\Omega),
\end{equation}
with $\mathbf{\Omega} =(\Omega_1,\Omega_2,\dots, \Omega_N)$. Thus any many-spin operator can be written as
\begin{equation}
    \hat A = \int\! d^N\Omega\, {\cal A}(\Om)\,  \hat{\mathbf{\Delta}}(\mathbf{\Omega}),\label{eq:A}
\end{equation}
where the integration $d^N\Omega=\prod_{j=1}^N d\Omega_j$ is over the angles $\theta_j,\phi_j$
\begin{equation}
    \int d\Omega = \frac{1}{2\pi}\int_0^{2\pi}\!\!\! d\phi \int_0^\pi \!\! d\theta\sin\theta.
\end{equation}
Note that with the above normalisation $\int d\Omega = 2$, which was chosen in order to have the Weyl symbol
of the unit matrix $\mathbf{1}$ to be the number $1$.
\begin{equation*}
    \mathbf{1} = \int\! d\Omega \, \hat\Delta(\Omega).
\end{equation*}
It is useful to decompose the phase-point operators in terms of Pauli matrices and spherical harmonics $\Y_l^m(\theta,\phi)$:
%
%
%
\begin{align}
    &\hat{\Delta}(\theta, \phi) = \sqrt{\pi}\Y_0^0(\theta,\phi) \, \mathbf{1}+\sqrt{\pi} \Y_1^0(\theta,\phi)\, \sigma_z \label{eq:Delta}\\
    &\quad - \sqrt{2\pi}\, \textrm{Re}\left[\Y_1^1(\theta,\phi)\right] \sigma_x
    - \sqrt{2\pi}\,  \textrm{Im}\left[\Y_1^1(\theta,\phi)\right] \sigma_y.\nonumber
\end{align}
Due to the orthogonality of the spherical harmonics only terms in ${\cal A}(\Omega)$ proportional to $Y_l^m$ with $l\le 1$ give a non-vanishing contribution.
Thus, although 
%
\begin{align}
& \tr\left\{\hat\Delta(\Omega)\hat\Delta(\Omega^\prime)\right\}  = 
    2\pi \Bigl(\Y_0^0(\Omega) \Y_0^0(\Omega^\prime) + \Y_1^0(\Omega) \Y_1^0(\Omega^\prime) \nonumber \\
    &\qquad \qquad+ \Y_1^{1*}(\Omega) \Y_1^{1}(\Omega^\prime) + \Y_1^{-1*}(\Omega) \Y_1^{-1}(\Omega^\prime)\Bigr) \label{eq:overcomplete}\\
&\enspace =  \delta(\Omega-\Omega^\prime) -2\pi\sum_{l>1}\sum_{m=-l}^l \Y_l^{m*}(\Omega) \Y_l^m(\Omega^\prime)\ne \delta(\Omega-\Omega^\prime), \nonumber 
\end{align}
where we have used the completeness relation of spherical harmonics, eq.~\eqref{eq:complete}, one can invert eq.~\eqref{eq:A} to obtain the
Weyl symbol ${\cal A}$
\begin{equation}
    {\cal A}(\Om) = \tr\left\{\hat{A}\, \Dm(\Om)\right\}.\label{eq:Weyl}
\end{equation}
Here the trace is over all $N$ spin degrees. 

One can also show directly that eq.~\eqref{eq:Weyl} is  correct for the case of a single-spin operator $\hat A = \alpha \mathbf{1} + \beta \sigma_x +\gamma \sigma_y +\lambda \sigma_z$, whose Weyl symbol according to~\eqref{eq:Weyl} reads:
\begin{align}
     {\cal A}  &= \alpha \sqrt{4\pi} \Y_0^0(\theta,\phi) + (\beta + i\gamma)  \sqrt{2\pi}\Y_1^{-1}(\theta,\phi) \nonumber\\
     &+ \lambda \sqrt{4\pi}  \Y_1^0(\theta,\phi) - (\beta - i\gamma)\sqrt{2\pi} \Y_1^{1}(\theta,\phi) ,\nonumber
\end{align}
and thus, using $\Y_1^{-1} = - \left(\Y_1^1\right)^*$ and the orthogonality of spherical harmonics,
\begin{eqnarray}
    \int \! \! d\Omega \, \, \Y_l^m(\Omega)^* \Y_{l^\prime}^{m^\prime}(\Omega) = \frac{1}{2\pi}\delta_{ll^\prime}\delta_{mm^\prime}\label{eq:ortho}
\end{eqnarray}
one finds 
\begin{align}
    \int d\Omega\,  {\cal A}(\Omega) \, \hat{\Delta}(\Omega) &= \alpha \mathbf{1}+\beta \sigma_x +\gamma \sigma_y+\lambda \sigma_z .\nonumber
\end{align}
%

\subsubsection{Density matrix and Wigner function}
   
Special importance has the Weyl symbol of the density operator $\rh$, which we will call Wigner function for spins
\begin{equation}
    \rh = \int\!\! d\Om\, \W(\Om)\, \Dm(\Om).
    \label{eq:wigner_rho_definition}
\end{equation}
Applying the inverse relation between Weyl symbol and operators, eq.~\eqref{eq:Weyl}, to the density operator yields an explicit expression for the
Wigner function
\begin{eqnarray}
    \W(\Om) = \tr\left\{\hat{\rho}\, \Dm(\Om)\right\} = \W^*(\Om),\label{eq:Wigner}
\end{eqnarray}
which is real, but not necessarily positive.

The normalization of the density matrix $\hat{\rho}$ results in the corresponding 
normalization of the Wigner function
\begin{equation}
    \int\! \!  d\Omega \, \W(\Omega) = 1.
\end{equation}

Since the phase-point operators $\hat\Delta(\Omega)$ are not positive semidefinite for all values of the angles $\theta$ and $\phi$, not all Wigner functions fulfilling the normalization conditions correspond to physical states. 

For a single spin, positivity of the density matrix implies the positivity of the determinant, which can be expressed as

\begin{eqnarray*}
    \det \hat \rho = \frac{1}{4} \Bigl\{ 1- 3\, 
\overline{\cos\theta}^2 - 3\vert\,  \overline{e^{-i\phi}\sin\theta}\, \vert^2\Bigr\}
\end{eqnarray*}
%
%
%
Thus, e.g. axial symmetric Wigner functions 
%
%
which do not have too large weight for $\theta$ angles close to the poles, represent true quantum states. This needs to be taken into account when doing approximations to the equations of motion.


\subsubsection{Expectation values}

With these definitions, expectation values of operators  can be expressed in terms of the Wigner function and the Weyl symbol of the operator:
\begin{eqnarray}
    \langle \hat A \rangle &=& \tr\{\hat A \rh\} = \int\!\! d\Om\, \W(\Om)\, \tr\{\hat A \Dm(\Om)\} \nonumber\\
    &= & \int\!\! d\Om\, \W(\Om)\, \cal{A}(\Om).\label{eq:expectation}
\end{eqnarray}
This expression looks very similar to a classical average of ${\cal A}(\Omega)$ weighted by the quasi-probability $\W(\Omega)$. $\W(\Omega)$, while being real, is in general, however, not positive and thus cannot be regarded as probability distribution.

\subsubsection{Reduced density matrix}

For some problems, the density matrix of a reduced subset of spins is needed. If the total system consisting of subsystems $A$ and $B$ is described by a density matrix $\hat \rho$, the reduced matrix
$\hat \rho_B = \textrm{Tr}_A\{\hat \rho\}$ has a Weyl symbol
\begin{equation}
    \hat \rho_B \, \, \longleftrightarrow \, \, \W_B(\Om) = \int_{A} \!\!\!d\Om_A\, \W(\Om),
\end{equation}
where $d\Omega_{A,B} = \prod_{j\in {A,B}} d\Omega_j$ is the set of angle variables corresponding to all spins in $A$ or $B$, respectively.

%
%

\subsection{gauge freedom}

The Weyl symbol of an operator is not unique, and so is the Wigner function.
Using the orthonormality relation of spherical harmonics, eq.~\eqref{eq:ortho}, one notices that one can add any term proportional to higher spherical harmonics $\Y_{l>1}^m(\theta,\phi)$ to $\W$ 
\begin{equation}
    \W \enspace \longrightarrow \enspace \W^\prime = \W+ \sum_{l>1}\sum_{m=-l}^l c_{lm}\Y_{l}^m(\theta,\phi)\label{eq:gauge}
\end{equation}
without changing the density operator $\hat \rho$, since
\begin{eqnarray}
    \int \!d\Omega \, \Y_{l>1}^m(\theta,\phi)\, \hat\Delta(\Omega) = 0.
\end{eqnarray}
Eq.~\eqref{eq:gauge} constitutes a \textit{gauge freedom}, which can be exploited to modify the phase-space representation of spin states.
In particular, as illustrated in the following subsections, it can be used to obtain positive definite Wigner functions for states, whose
representation in terms of lowest-order spherical harmonics is not. This is an important advantage of the continuous phase-space representation for two reasons: (i) The simulation of many-body dynamics by \acp{sde}
is only possible if the initial state can be sampled, i.e. has a positive Wigner distribution. (ii) As a Fokker-Planck equation always maps positive Wigner functions to positive Wigner functions,
an accurate semiclassical approximation of the quantum dynamics can only be expected if both initial and final quantum states have a positive Wigner function.
For discrete representations such as the \ac{dtwa} this is in general not the case. 

Note, that the transformation, eq.~\eqref{eq:gauge}, can in principle also be applied to the Weyl symbol ${\cal A}$ of an observable. However, expression~\eqref{eq:expectation} for expectation values  only remains correct, if the gauge transformation is applied either to the Wigner function or to the Weyl symbol of the observable, but not to both. For this reason we will apply gauge transformations only to the Wigner function. It should then be further noted
that for the same reason gauge transformations cannot be applied to expressions that are nonlinear in the density matrix, such as the Renyi or von-Neumann entropies.

\subsection{Weyl symbols of spin operators and Wigner functions of spin states}

\begin{figure}[h]
\begin{center}
\includegraphics[width=0.3\textwidth]{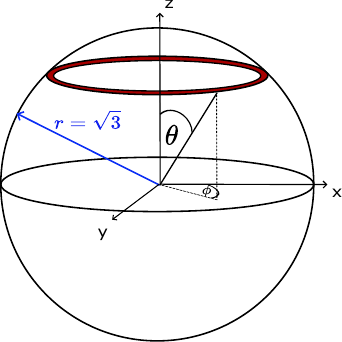}
\end{center}
\vspace{-5mm}
\caption{Continuous phase-space representation of $\vert\uparrow\rangle$ state} 
\label{fig:Bloch}
\end{figure}

To find Weyl symbols for spin operators, respectively the Wigner function of a density matrix we need to
evaluate the trace of the product of the corresponding operator with the phase point operator $\hat \Delta (\Omega)$. Since all many-particle spin states and observables can be written as Kronecker products of $2\times 2$ matrices for individual spins, it is sufficient for this to consider a single spin. The trace can be straightforwardly calculated using the expansion of $\hat \Delta(\Omega)$ in terms of Pauli matrices, eq.~\eqref{eq:Delta}. From this we find the following Weyl symbols
\begin{eqnarray}
\hat A \quad && \quad {\cal A}\nonumber\\
\mathbf{1} \, \, \enspace &\leftrightarrow & \quad 1,\nonumber \\
    \hat \sigma_x \enspace &\leftrightarrow & \enspace \sqrt{3}\,\sin\theta\cos\phi,\\
    \hat \sigma_y \enspace &\leftrightarrow & \enspace \sqrt{3}\,\sin\theta\sin\phi,\nonumber \\
    \hat \sigma_z \enspace &\leftrightarrow & \enspace \sqrt{3}\,\cos\theta. \nonumber
\end{eqnarray}
In a similar way we directly obtain the Wigner function of states. E.g. spin-up $\rho_\uparrow = \vert \!\uparrow\rangle\langle\uparrow\!\vert = \frac{1}{2}(1+\hat \sigma_z)$ and spin-down states $\rho_\downarrow = \vert \!\downarrow\rangle\langle\downarrow\!\vert = \frac{1}{2}(1-\hat \sigma_z)$ correspond to the Wigner functions
\begin{eqnarray}
\hat \rho  \quad\,  && \qquad {\W}\nonumber\\
\vert \!\uparrow\rangle\langle\uparrow\!\vert  \enspace &\leftrightarrow & \enspace \frac{1}{2}\Bigl(1 +\sqrt{3}\cos\theta\Bigr),\\
\vert \!\downarrow\rangle\langle\downarrow\!\vert  \enspace &\leftrightarrow & \enspace \frac{1}{2}\Bigl(1 -\sqrt{3}\cos\theta\Bigr). \nonumber
\end{eqnarray}
These expressions have the disadvantage that they are not positive definite and thus cannot be interpreted as a probability distribution. However, we can make use of the gauge freedom, eq.~\eqref{eq:gauge}, to find equivalent Wigner functions, that are positive. E.g. the spin-up state can be represented as
\begin{equation}
    \W_{\vert\uparrow\rangle\langle\uparrow\vert}^\prime = \frac{1}{\sin\theta}
    \delta\Bigl(\theta-\arccos\left(\frac{1}{\sqrt{3}}\right)\Bigr),
\end{equation}
where $\cos\theta$ is fixed to $1/\sqrt{3}$ and $\phi$ is equally distributed.
One easily verifies using eq.~\eqref{eq:Delta}
that
\begin{equation*}
    \int d\Omega\, \W_{\vert\uparrow\rangle\langle\uparrow\vert}^\prime (\Omega)\, \hat \Delta(\Omega) = \frac{1}{2}\Bigl(1+\hat \sigma_z\Bigr) = \vert\! \uparrow\rangle\langle \uparrow\!\vert.
\end{equation*}
$\W_{\vert\uparrow\rangle\langle\uparrow\vert}^\prime$ is illustrated in Fig.~\ref{fig:Bloch}. 
%
%
Here one recognizes the importance of the increased radius $\sqrt{3}$ of the ''Bloch'' sphere in the phase space mapping: The spin-up state does not correspond to a vector pointing to the north pole, but the polar angle is fixed at the non-zero value $\theta=\arccos(1/\sqrt{3})$. As a consequence the $x$ and $y$ projections of the spin of that state have finite fluctuations as required by Heisenberg's uncertainty relation.
\begin{equation}
    \langle \Delta \sigma_x^2\rangle = 3 \int\! d\Omega\,  
    \W_{\vert\uparrow\rangle\langle\uparrow\vert}^\prime(\Omega) \sin^2\theta\, \cos^2\phi = 1.
\end{equation}


\subsection{Wigner function of entangled spin states}
Non-Gaussian, non-classical states of bosonic fields are often characterized by non-positive definite Wigner functions.  Surprisingly, this is different in the case of spins. Here, even for maximally entangled states, such as Bell states, there exist gauges in which the Wigner function is positive. This has an important consequence. As we will show for a specific example later, the \ac{twa} is, in certain cases, capable of describing processes in which entanglement is generated.
One finds that the following expressions are proper Wigner functions of the four Bell states
$\vert \Psi_\pm\rangle$ and $\vert \Phi_\pm\rangle$:
\begin{eqnarray}
    \W_{\Phi_+} &=& \frac{\pi}{\sin\theta_1\sin\theta_2}
    \Bigl[\delta(\theta_1-\theta_\uparrow) \delta(\theta_2-\theta_\uparrow) \nonumber\\
   &&\qquad+ \delta(\theta_1-\theta_\downarrow) \delta(\theta_2-\theta_\downarrow)
    \Bigr]\delta(\phi_1+\phi_2),
    \nonumber\\
  \W_{\Phi_-} &=& \frac{\pi}{\sin\theta_1\sin\theta_2}
    \Bigl[\delta(\theta_1-\theta_\uparrow) \delta(\theta_2-\theta_\uparrow) \label{eq:Bell-states}\\
   &&\qquad + \delta(\theta_1-\theta_\downarrow) \delta(\theta_2-\theta_\downarrow)
    \Bigr]\delta(\phi_1+\phi_2-\pi),
  \nonumber  \\
\W_{\Psi_+} &=& \frac{\pi}{\sin\theta_1\sin\theta_2}
    \Bigl[\delta(\theta_1-\theta_\uparrow) \delta(\theta_2-\theta_\downarrow) \nonumber\\
   &&\qquad+ \delta(\theta_1-\theta_\downarrow) \delta(\theta_2-\theta_\uparrow)
    \Bigr]\delta(\phi_1-\phi_2),
    \nonumber\\
\W_{\Psi_-} &=& \frac{\pi}{\sin\theta_1\sin\theta_2}
    \Bigl[\delta(\theta_1-\theta_\uparrow) \delta(\theta_2-\theta_\downarrow) \nonumber\\
   &&\qquad+ \delta(\theta_1-\theta_\downarrow) \delta(\theta_2-\theta_\uparrow)
    \Bigr]\delta(
    \phi_1-\phi_2-\pi),
    \nonumber
\end{eqnarray}
where the Delta functions in $\phi$ are all modulo $2\pi$ and $\theta_{\uparrow/\downarrow} = \arccos\bigl(\pm 1/\sqrt{3}\bigr)$. All of these expressions are positive!
This is in contrast to Wootters 4-point discrete representation, for which some Wigner coefficients of Bell states are negative, and these states cannot be the result of a SDE simulation (i.e. a Fokker-Planck dynamics) in this discrete representation.

\section{Equations of motion of $\W(\Omega)$ and Fokker-Planck approximation} \label{sec:FPE}

The importance of a phase-space representation of spins results from the fact that it provides a systematic approach to derive approximate equations of motion that are amenable to numerically inexpensive simulation techniques, which we will discuss in the following in detail.

\subsection{General procedure}

When discussing open spin systems we consider an ensemble of $N$ spins coupled to
reservoirs. For Markovian reservoirs the density operator obeys a Lindblad master equation
%
%
%
%
%
%
%
%
\begin{equation}
    \partial_t \hat \rho(t) = - i[H_s,\hat \rho(t)] + {\cal L}_\textrm{diss}\hat \rho(t) = {\cal L} \hat  \rho(t),\label{eq:Lindblad}
\end{equation}
where ${\cal L}$ is a Lindblad superoperator including the action of $H_s$, denoting the system Hamiltonian, and dissipative couplings, ${\cal L}_\textrm{diss}$. 
%
%
In order to translate the Lindblad equation, eq.~\eqref{eq:Lindblad}, into an \ac{eom} of the Wigner function one simply has to insert eq.~\eqref{eq:wigner_rho_definition} on both sides. 
This then leads to 
\begin{align}
    &\partial_t\rh(t) = \int\!d\Om\,  \partial_t \W(\Om,t) \Dm(\Om) \nonumber\\
    &= \int \!d\Om\,  \W(\Om,t) [{\cal L} \Dm(\Om)]. 
\end{align}
The action of ${\cal L}$ on the phase point operators corresponds to evaluating products of Pauli spin operators and phase point operators.
Such products can, however, be expressed in terms of a complete set of operators in the two-dimensional spin space, formed by the four operators, e.g.
\begin{equation}
    \hat \Delta,\enspace \partial_\theta  \hat \Delta, \enspace  \partial_\phi  \hat \Delta, \enspace  \partial^2_\phi  \hat \Delta,\label{eq:set}
\end{equation}
i.e. by the phase point operator $\hat\Delta(\Omega)$ itself and its partial derivatives with respect to $\theta$ and $\phi$.
It should be noted that this choice for a set of linear independent operators is not unique and there may be other choices better suited for certain problems.
In particular, the second derivative with respect to $\phi$ could be replaced with the second derivative with respect to $\theta$.

By partial integration, these derivatives can subsequently be transferred to the Wigner function in eq.~\eqref{eq:EOM}. 
Here one has to pay attention, however, to the curved phase space, where the infinitesimal volume elements, $d\Omega = d\theta d\phi \, \sin\theta$, are not constant, which complicates the partial integration. To avoid this, it is
more convenient to introduce the \textit{flattened} Wigner function
\begin{equation}
    \chi(\Omega) ={\cal W}(\Omega)\, \pi\, \sin\theta,
\end{equation}
along with the \textit{flattened} integral measure
\begin{equation}
  d\tilde{\Omega} = \frac{ d\Omega  }{\pi \sin\theta},\qquad \int \! d\tilde\Omega = 
  \frac{1}{2\pi^2} \int_0^\pi \!\!\! d\theta  \int_0^{2\pi} \!\!\!\! d\phi = 1
\end{equation}
such that
\begin{eqnarray}
     \langle \hat A\rangle = \int\! d\Omega\, \W(\Omega)\, {\cal A}(\Omega)
     = \int\! d\tilde\Omega\, \chi(\Omega)\, {\cal A}(\Omega),
\end{eqnarray}
and
\begin{align}
     &\partial_t\rh(t) = \int\!\!d \Omt\,\,   \partial_t \chi(\Om,t) \Dm(\Om) = \int \!\!d\Omt\,   \chi(\Om,t) [{\cal L} \Dm(\Om)]   \nonumber\\
    &=\! \int\! \!d\Omt\,   \chi(\Om,t) [{\cal D} \Dm(\Om)] 
    = \! \int \!\!d\Omt\,   [\tilde {\cal D} \chi(\Om,t)]\Dm(\Om). 
  \label{eq:EOM} 
\end{align}
Here ${\cal D}$ denotes the differential operator resulting from rewriting the action of the Lindbladian ${\cal L}$ on $\Dm$ in terms of derivaties, eq.~\eqref{eq:set}, and $\tilde {\cal D}$ is the corresponding differential operator after partial integration. 

For products of multiple spin operators, this procedure can be repeated, eventually leading to a partial differential equation for $\chi(\Om,t)$. Since conservation of probability demands

\begin{align}
    \int d\Omt\, \chi(\Om,t) = \textrm{const}_t\nonumber
\end{align}
only terms containing derivatives of $\chi(\Om,t)$ occur, and the \ac{eom} reads 
\begin{align}
    &\frac{\partial}{\partial t} \chi(\Om,t) 
     = \tilde{\cal D}\chi(\Om,t) \nonumber \\
   & = \sum_j \frac{\partial}{\partial x_j} A_j \chi(\Om,t)
    + \sum_{jl}\frac{\partial^2}{\partial x_j \partial x_l} D_{jl} \chi(\Om,t) \label{eq:W-EOM}\\
    & \qquad + \sum_{jlm}\frac{\partial^3}{\partial x_j \partial x_l \partial x_m} G_{jlm} \chi(\Om,t) +\cdots \nonumber
\end{align}

where the $x_j\in\{\theta_j,\phi_j\}$ correspond to the angle variables of the spins. 
Furthermore, since $\chi(\Om,t)$ is real, the diffusion matrix $D_{jl}$ is real and symmetric, i.e. 
\begin{equation}
    D_{jl} = D_{jl}^* = D_{lj},
\end{equation}
but not necessarily positive, i.e. it may not correspond to a diffusion process in phase space.

\subsection{Direct correspondence rules}

In the \ac{eom}, eq.~\eqref{eq:EOM}, we have to evaluate the action of the Lindbladian on the phase-space operator, ${\cal L}\Dm$.  This contains
expressions
of the type $[H,\Dm]$ and $\hat L \Dm \hat{L}^\dagger - \frac{1}{2}\{\hat{L}^\dagger \hat{L},\Dm\}$. In the following we will outline a general procedure of how to translate commutators with spin Hamiltonians and the action of Lindbladians into phase space. These rules are termed \textit{correspondence rules}.
Since the Hamiltonian $H$ and the Lindblad operators
$\hat L$ can be written in terms of products of single-spin operators, it is sufficient to consider the products of a single spin operator with 
$\hat \Delta(\Omega)$. 
These products will be expressed in terms of derivatives 
of $ \hat \Delta(\Omega)$, which then leads to a set of \textit{direct correspondence rules}. The derivatives will then be transferred to the Wigner function in eq.~\eqref{eq:EOM} by partial integration which eventually leads to the partial differential equation ~\eqref{eq:W-EOM}. 
%
%
%
%
The procedure can straightforwardly be iterated to products with multiple spin operators.

\begin{table}[!ht]
    \centering 
    \caption{Correspondence rules for the phase point operator acting with one Pauli matrix. (Notation: $\hat\sigma_\mu \hat\Delta$ or $\hat\Delta\hat\sigma_\mu$ (upper  or lower sign) is mapped to $\beta_0\hat{\Delta}+\beta_1\,\partial_\theta\hat{\Delta}+\beta_2\,\partial_\phi\hat{\Delta}+\beta_3\,\partial_\phi^2\hat{\Delta}$ or $\gamma_0\hat{\Delta}+\gamma_1\,\partial_\theta\hat{\Delta}+\gamma_2\,\partial_\phi\hat{\Delta}+\gamma_3\,\partial_\theta^2\hat{\Delta}$)}
    \label{tab:table-1}
    \begin{tabular}{c|c|c}
        ~ & 
        $\left[\beta_0, \beta_1, \beta_2, \beta_3 \right]^\top$ &
        $\left[\gamma_0, \gamma_1, \gamma_2, \gamma_3 \right]^\top$
        \\
        \hline
        $\begin{matrix}
            \hat\sigma_x\hat{\Delta} \\ \hat{\Delta}\hat\sigma_x
        \end{matrix}$
        & 
        $\begin{matrix}
            \sqrt{3}\sin\theta\cos\phi \\ \sqrt{3}\cos\theta\cos\phi \mp i\sin\phi \\ -\frac{\csc\theta}{\sqrt{3}}\sin\phi \mp i\cot\theta\cos\phi \\ \frac{2}{\sqrt{3}}\csc\theta\cos\phi 
        \end{matrix}$ &
        $\begin{matrix}
            \sqrt{3}\sin\theta\cos\phi \\ \frac{1}{\sqrt{3}}\cos\theta\cos\phi \mp i\sin\phi \\ -\frac{\csc\theta}{\sqrt{3}}\sin\phi \mp i\cot\theta\cos\phi \\ \frac{2}{\sqrt{3}}\sin\theta\cos\phi  
        \end{matrix}$
        \\
        \hline
        $\begin{matrix}
            \hat\sigma_y\hat{\Delta} \\ \hat{\Delta}\hat\sigma_y
        \end{matrix}$ &
        $\begin{matrix}
            \sqrt{3}\sin\theta\sin\phi \\ \sqrt{3}\cos\theta\sin\phi \pm i\cos\phi \\ \frac{\csc\theta}{\sqrt{3}}\cos\phi \mp i\cot\theta\sin\phi \\ \frac{2}{\sqrt{3}}\csc\theta\sin\phi
        \end{matrix}$ &
        $\begin{matrix}
            \sqrt{3}\sin\theta\sin\phi \\ \frac{1}{\sqrt{3}}\cos\theta\sin\phi \pm i\cos\phi \\ \frac{\csc\theta}{\sqrt{3}}\cos\phi \mp i\cot\theta\sin\phi \\ \frac{2}{\sqrt{3}}\sin\theta\sin\phi
        \end{matrix}$
        \\ 
        \hline
        $\begin{matrix}
            \hat\sigma_z\hat{\Delta} \\ \hat{\Delta}\hat\sigma_z
        \end{matrix}$ &
        $\begin{matrix}
            \sqrt{3}\cos\theta \\ \frac{1}{\sqrt{3}}\left(2\csc\theta-3\sin\theta\right) \\ \pm i \\ \frac{2}{\sqrt{3}}\csc\theta\cot\theta
        \end{matrix}$ &
        $\begin{matrix}
            \sqrt{3}\cos\theta \\ -\frac{1}{\sqrt{3}}\sin\theta \\ \pm i \\ \frac{2}{\sqrt{3}}\cos\theta
        \end{matrix}$
        \\ 
        \hline
        $\hat\sigma_\pm\hat{\Delta}$ & 
        $\begin{matrix}
            \frac{\sqrt{3}}{2}\sin\theta\,\mathrm{e}^{\pm i\phi} \\ \frac{1}{2}\left(\sqrt{3}\cos\theta \mp 1\right)\,\mathrm{e}^{\pm i\phi} \\ -i\frac{\csc\theta}{2\sqrt{3}}\left(\sqrt{3}\cos\theta \mp 1\right)\mathrm{e}^{\pm i\phi} \\ \frac{1}{\sqrt{3}}\csc\theta\,\mathrm{e}^{\pm i\phi}
        \end{matrix}$ & 
        $\begin{matrix}
            \frac{\sqrt{3}}{2}\sin\theta\,\mathrm{e}^{\pm i\phi} \\ \frac{1}{2}\left(\frac{\cos\theta}{\sqrt{3}} \mp 1\right)\,\mathrm{e}^{\pm i\phi} \\ -i\frac{\csc\theta}{2\sqrt{3}}\left(\sqrt{3}\cos\theta \mp 1\right)\mathrm{e}^{\pm i\phi} \\ \frac{1}{\sqrt{3}}\sin\theta\,\mathrm{e}^{\pm i\phi}
        \end{matrix}$
        \\ 
        \hline
        $\hat{\Delta}\hat\sigma_\pm$ & 
        $\begin{matrix}
            \frac{\sqrt{3}}{2}\sin\theta\,\mathrm{e}^{\pm i\phi} \\ \frac{1}{2}\left(\sqrt{3}\cos\theta \pm 1\right)\mathrm{e}^{\pm i\phi} \\ i\frac{\csc\theta}{2\sqrt{3}}\left(\sqrt{3}\cos\theta \pm 1\right)\mathrm{e}^{\pm i\phi} \\ \frac{1}{\sqrt{3}}\csc\theta\,\mathrm{e}^{\pm i\phi} \end{matrix}$ &
        $\begin{matrix}
            \frac{\sqrt{3}}{2}\sin\theta\,\mathrm{e}^{\pm i\phi} \\ \frac{1}{2}\left(\frac{\cos\theta}{\sqrt{3}} \pm 1\right)\,\mathrm{e}^{\pm i\phi} \\ i\frac{\csc\theta}{2\sqrt{3}}\left(\sqrt{3}\cos\theta \pm 1\right)\mathrm{e}^{\pm i\phi} \\ \frac{1}{\sqrt{3}}\sin\theta\,\mathrm{e}^{\pm i\phi}
        \end{matrix}$
        \\
    \end{tabular}
\end{table}

The derivation of the direct correspondence rules is tedious but straightforward.
For example 
we can evaluate
%
\begin{eqnarray*}
   && \sigma_\pm \hat{\Delta} = \Bigl(\frac{1}{2}\mp\frac{\sqrt{3}}{2}\cos\theta\Bigr) \sigma_\pm + \frac{\sqrt{3}}{4} e^{\pm i\phi} \sin\theta(\mathbf{1}\pm \sigma_z)\\
    & &\enspace = \frac{\sqrt{3}}{2} e^{\pm i\phi} \sin\theta\,\hat{\Delta} \mp \frac{1}{2}\left( 1 \mp \sqrt{3} \cos\theta \right)\mathrm{e}^{\pm i\phi}\,\partial_\theta\hat{\Delta} + \\
    &&\quad \pm \frac{i}{2}\frac{\csc\theta}{\sqrt{3}}\left( 1 \mp \sqrt{3}\cos\theta \right) \mathrm{e}^{\pm i\phi}\,\partial_\phi\hat{\Delta} + \frac{\csc\theta}{\sqrt{3}}\mathrm{e}^{\pm i\phi}\,\partial^2_\phi\hat{\Delta}.
\end{eqnarray*}
In table~\ref{tab:table-1} we have listed the correspondence rules for all products with a single Pauli matrix and in table~\ref{tab:table-2} we have given the most relevant expressions for products of two Pauli matrices. Two variants of the direct correspondence rules are listed, those including $\partial^2_\phi \hat \Delta$ and those including $\partial^2_\theta \hat \Delta$.

\begin{table}[ht]
    \centering
    \caption{Correspondence rules for the phase point operator acting with two Pauli matrices. (Notation: $\hat\sigma_\mu \hat \Delta \hat \sigma_\nu$ etc. $\longleftrightarrow\beta_0\hat{\Delta}+\beta_1\,\partial_\theta\hat{\Delta}+\beta_2\,\partial_\phi\hat{\Delta}+\beta_3\,\partial_\phi^2\hat{\Delta}$ or $\gamma_0\hat{\Delta}+\gamma_1\,\partial_\theta\hat{\Delta}+\gamma_2\,\partial_\phi\hat{\Delta}+\gamma_3\,\partial_\theta^2\hat{\Delta}$)}
    \label{tab:table-2}
    \begin{tabular}{c|c|c}
        ~ & 
        $\left[\beta_0, \beta_1, \beta_2, \beta_3 \right]^T$ &
        $\left[\gamma_0, \gamma_1, \gamma_2, \gamma_3 \right]^\top$
        \\
        \hline
        $\sigma_z\hat{\Delta}\sigma_z$ & 
        $\begin{matrix}
            1 \\ 0 \\ 0 \\ 2
        \end{matrix}$ &
        $\begin{matrix}
            1 \\ -2\sin\theta\cos\theta \\ 0 \\ 2\sin^2\theta
        \end{matrix}$ 
        \\ 
        \hline
        $\sigma_\pm\hat{\Delta}\sigma_\mp$ & 
        $\begin{matrix}
            \frac{1}{2}\left(1\mp\sqrt{3}\cos\theta\right) \\ \mp\frac{\csc\theta}{2\sqrt{3}}\left(\sqrt{3}\cos\theta \mp 1\right)^2 \\ 0 \\ \csc^2\theta-\frac{1}{2}\mp\frac{2}{\sqrt{3}}\csc\theta\cot\theta
        \end{matrix}$ &
        $\begin{matrix}
            \frac{1}{2}\left(1 \mp \sqrt{3}\cos\theta\right) \\ \frac{\sin\theta}{2\sqrt{3}}\left(\sqrt{3}\cos\theta \mp 1\right) \\ 0 \\ 1-\frac{1}{2}\sin^2\theta\mp\frac{2}{\sqrt{3}}\cos\theta
        \end{matrix}$ 
        \\
        \hline
        $\sigma_\pm\sigma_\mp\hat{\Delta}$ & 
        $\begin{matrix}
            \frac{1}{2}\left(1\pm\sqrt{3}\cos\theta\right) \\ \pm\frac{1}{2\sqrt{3}}\left(2\csc\theta-3\sin\theta\right) \\ \pm\frac{i}{2} \\ \pm\frac{1}{\sqrt{3}}\csc\theta\cot\theta
        \end{matrix}$ &
        $\begin{matrix}
            \frac{1}{2}\left(1\pm\sqrt{3}\cos\theta\right) \\ \mp \frac{1}{2\sqrt{3}}\sin\theta\\ \pm\frac{i}{2} \\ \pm\frac{1}{\sqrt{3}}\cos\theta
        \end{matrix}$ 
        \\ 
        \hline
        $\hat{\Delta}\sigma_\pm\sigma_\mp$ & 
        $\begin{matrix}
            \frac{1}{2}\left(1\pm\sqrt{3}\cos\theta\right) \\ \pm\frac{1}{2\sqrt{3}}\left(2\csc\theta-3\sin\theta\right) \\ \mp\frac{i}{2} \\ \pm\frac{1}{\sqrt{3}}\csc\theta\cot\theta
        \end{matrix}$ &
        $\begin{matrix}
            \frac{1}{2}\left(1\pm\sqrt{3}\cos\theta\right) \\ \mp \frac{1}{2\sqrt{3}}\sin\theta \\ \mp\frac{i}{2} \\ \pm\frac{1}{\sqrt{3}}\cos\theta
        \end{matrix}$ 
        \\        
    \end{tabular}
\end{table}

\

\subsection{Approximate correspondence rules for collective couplings}\label{sect:approximate}

For  problems with collective interactions there is an alternative and simpler way to construct approximate \textit{collective correspondence rules}. For cooperative interactions, the most relevant physical quantities are collective spin operators of the type
\begin{equation}
    \boldsymbol{\hat{S}}(c_n) = \sum_{n=1}^N c_n \bold{\hat{\sigma}}_n,
    \label{eq:coll_spin_op}
\end{equation}
where the weights $c_n$ determine the degree of cooperativity. This is the case, e.g., for an ensemble of spins interacting collectively with the quantized electromagnetic field, where the distances between the spins are comparable to the wavelength $\lambda_e$ of the dipole interaction, which leads to effects such as superradiance.


 As shown in~\cite{mink2023collective} the contribution
 of the collective operators~\eqref{eq:coll_spin_op} to the EOM of the Wigner function can be approximated as
\begin{align}
    \boldsymbol{\hat{S}}(c_n) \, \hat{\rho}  \longrightarrow & \, \boldsymbol{\mathcal{S}}(c_n) \, \W( \boldsymbol{\Omega}) \nonumber \\ \quad =& \sum_{n=1}^N c_n \cdot \bigl(\boldsymbol{s}_n + \boldsymbol{L}_n\bigr) \W(\boldsymbol{\Omega} ) 
    \label{coll_weyl_truncated}
\end{align}
and
\begin{align}
    \hat{\rho} \boldsymbol{\hat{S}}(c_n)\longrightarrow \sum_{n=1}^N c_n \cdot \bigl(\boldsymbol{s}_n - \boldsymbol{L}_n\bigr) \W(\boldsymbol{\Omega} )
\end{align}
with the differential operators
\begin{align}
    \boldsymbol{L}_n f = i \nabla_{\theta_n}
    \begin{pmatrix}
        + \sin{\phi_n} \\ - \cos{\phi_n} \\ 0
    \end{pmatrix} f
    + i \nabla_{\phi_n}
    \begin{pmatrix}
        \cot{\theta_n \cos{\phi_n}} \\ \cot{\theta_n} \sin{\phi_n} \\ -1
    \end{pmatrix} f,
    \label{coll_angular_momentum}
\end{align}
where $\boldsymbol{s}_n$ is the c-number vector from equation~\eqref{eq:CartesianSpin} and $\boldsymbol{\mathcal{S}}$ is the Weyl symbol of the collective operator. The approximation assumes that all relevant states of the system can be expressed in terms of collective excitations 
$\hat S^+ \vert 0\rangle $, with $\vert 0\rangle = \vert \!\!\downarrow\downarrow\dots \downarrow\rangle $ and $\sum_n \vert c_n\vert^2$ is large.
In these approximate correspondence rules the effect of the collective spin operators is split up in two parts, single-spin terms in phase space $\boldsymbol{s}_n$ and the angular momentum $\boldsymbol{L}_n$. 

The collective correspondence rules work very well for example for effects like superradiance~\cite{tebbenjohannsPRA2024,spahn2026motioninduceddirectionalitycollectiveemission} and, as will be discussed later, can even describe dissipative entanglement generation, but are not adequate when subradiant states become relevant~\cite{mink2023collective}. 
\par

%
%

\subsection{Fokker-Planck approximation and stochastic simulations}

\subsubsection{Fokker-Planck equation}

Solving the EOM of the Wigner function, eq.~\eqref{eq:W-EOM}, which are high-order partial differential equations is in general no simplification to the full quantum problem. However,
under certain conditions eq.~\eqref{eq:W-EOM} can be approximated by a $2N$-dimensional \textit{Fokker-Planck equation} (FPE). 
\begin{eqnarray}
    \frac{\partial}{\partial t} \chi(\Om,t)  &=& -\sum_j \frac{\partial}{\partial x_j} A_j(\Om) \chi(\Om,t) \label{eq:FPE}\\
    &&+ \frac{1}{2} \sum_{jl} \frac{\partial^2}{\partial x_j \partial x_l} D_{jl}(\Om) \chi(\Om,t),\nonumber
\end{eqnarray}
where the coefficient matrix of second-order derivatives is positive semi-definite and can thus be decomposed as $\mathbf{D}(\Om) = \mathbf{B}(\Om) \cdot\mathbf{B}(\Om)^\top$ and 
with the initial condition $\chi(\Om,t=0)=\chi_0(\Om) = \W_0(\Om) \prod_j\pi \sin\theta_j$ corresponding to the initial density operator.

Using the direct correspondence rules, one can show that   for a general \textit{unitary} time evolution the diffusion matrix, if existent, is  non-positive, and thus must be neglected in eq.~\eqref{eq:FPE}.  

For a general single-spin Hamiltonian $\hat H = a_0\hat\sigma_0 + a_1 \hat\sigma_x + a_2\hat\sigma_y + a_3\hat\sigma_z$ with arbitrary coefficients, the commutator with the phase-point operator $\hat\Delta$ leads to
\begin{align}
    \label{eq:General-Commutator}
    -i\left[\hat H, \hat\Delta \right] &= \sum_{i,j,k = 1}^3 \epsilon_{ijk}a_i s_j \hat\sigma_k \\
    &=
    \sqrt{3}
    \begin{pmatrix}
        \sin\theta\sin\phi \hat\sigma_z - \cos\theta\hat\sigma_y \\
        \cos\theta\hat\sigma_x-\sin\theta\cos\phi\hat\sigma_z \\
        \sin\theta\cos\phi\hat\sigma_y-\sin\theta\sin\phi\hat\sigma_x
    \end{pmatrix}
    \cdot \vec{a},
    \nonumber
\end{align}
where $\epsilon_{ijk}$ is the Levi-Civita symbol,  $\vec{a}=(a_1,a_2,a_3)$, and $\vec{s}$ is given in  eq.~\eqref{eq:CartesianSpin}. Now, each element of the left vector can be expressed in terms of $\partial_\theta \hat\Delta$ and $\partial_\phi \hat\Delta$, for instance
\begin{eqnarray*}
    &\sqrt{3}&\sin\theta\sin\phi\hat\sigma_z -\sqrt{3}\cos\theta\hat\sigma_y \\
    &=& -2\sin\phi\partial_\theta \hat\Delta- 2\cot\theta\cos\phi\partial_\phi \hat\Delta.
\end{eqnarray*}
Thus, as long as $\partial_\theta \hat\Delta$ and $\partial_\phi \hat\Delta$ are both used in the basis of direct correspondence rules, which is the most sensible approach, there is no contribution to the diffusion.

We now show that for general spin-spin interactions the diffusion matrix in eq.~\eqref{eq:FPE} is an indefinite matrix, more specifically, $\tr\left\{D_{jl}\right\}=0$. We first investigate the possible contributions to the diagonal of $D_{jl}$. One option is that the spin indices in the Hamiltonian are the same ($i=j$). In this case, however, the spin products map to the single-spin case which gives no contribution to the diffusion. Another option could be the second order derivatives $\partial^2_{\phi_i}\D$ or $\partial^2_{\theta_i}\D$ in the direct correspondence rules. While the product of first order derivatives yields only off-diagonal elements, second order derivatives in the same variable always vanish due to the commutator. Consequently $\tr\left\{D_{jl}\right\}=0$ and thus the diffusion matrix must have both positive and negative eigenvalues.
Positive diffusion terms only arise from dissipative processes.

\subsubsection{Mapping to Ito stochastic differential equations}

The solution of the FPE,~\eqref{eq:FPE}, can be written in terms of the conditional probability $p(\Om,t;\Om^\prime,0)$
\begin{align}
    \chi(\Om,t) &= \int\!d\Omt^\prime\, p(\Om,t; \Om^\prime,0) \, \chi_0(\Om^\prime).
\end{align}
$p(\Om,t;\Om^\prime,0)$ fulfills the same Fokker-Planck \ac{eom} as $\chi$, however, with the initial condition
\begin{align}
    p(\Om,0; \Om^\prime,0) = \delta(\Om-\Om^\prime).
\end{align}
Thus any expectation value can be calculated in phase space via
\begin{equation}
     \langle \hat A \rangle = \int\!\! d\Omt\!\int\!\! d\Omt^\prime \, {\cal A}(\Om)\, p(\Om,t; \Om^\prime,0) \, \chi_0(\Omega^\prime).
\end{equation}
The time evolution of the conditional probability function under a FPE can be efficiently simulated by a set of coupled Ito stochastic differential equations (SDE) for stochastic variables $x_j(t)\in\{ \phi_j(t),\theta_j(t)\}$ with initial conditions $\phi_j(0)=\phi_j^\prime$ and $\theta_j(0)=\theta_j^\prime$ sampled from $\W_0(\Om^\prime)$:
\begin{eqnarray}
    d x_j(t) = A_j[\Om(t)] \, dt + \sum_l B_{jl}[\Om(t)]\,  d W_l.
\label{eq:SDE}
\end{eqnarray}
Here $d W_l$ is a differential Wiener process. This provides an efficient way of (approximately) calculating integrals of the type
\begin{eqnarray}
  \langle \hat f\rangle = \int\! d\Omt\, f(\Om) \, \chi(\Om,t) 
     = \overline{f(\Om(t))}
\end{eqnarray}
as stochastic average, 
if $\chi_0$ is positive and can be interpreted as probability distribution.
This is done by the following
procedure:
\begin{itemize}
    \item[(i)] choose an initial value $\Omega^\prime$ for all spin variables 
    with probability weight $\chi_0(\Om^\prime)$.
    \item[(ii)] solve the Ito \acp{sde}~\eqref{eq:SDE} to obtain $\Omega(t)$ with these initial values
    \item[(iii)] evaluate the Weyl symbol of $\hat f$: $f(\Omega(t))$ 
    \item[(iv)] average over many trajectories (if the \acp{sde} have noise terms) and over the initial distribution $\chi_0(\Om^\prime)$.
\end{itemize}
Alternatively 
\begin{itemize}
    \item[(i)] choose random initial values $\Omega^\prime$ uniformly distributed on the unit sphere
     \item[(ii)] solve the Ito \acp{sde}~\eqref{eq:SDE} to obtain $\Omega(t)$ with these initial values
    \item[(iii)] calculate $f(\Omega(t))\chi_0(\Omega^\prime)$ 
    \item[(iv)] average over many trajectories (if the \acp{sde} have noise terms) and the initial flat distribution of $\Omega^\prime$.
\end{itemize}

\section{Stochastic differential equations for specific problems} \label{sec:SDEs}

In the following, we will explicitly give the \acp{sde} for important interaction Hamiltonians and Liouvillians. We will also illustrate the broad scope of applications of the \ac{twa} method and show its accuracy for some simple benchmark problems. 

\subsection{Single spin dynamics}

\begin{figure}[h!]
\begin{center}
\includegraphics[width=0.48\textwidth]{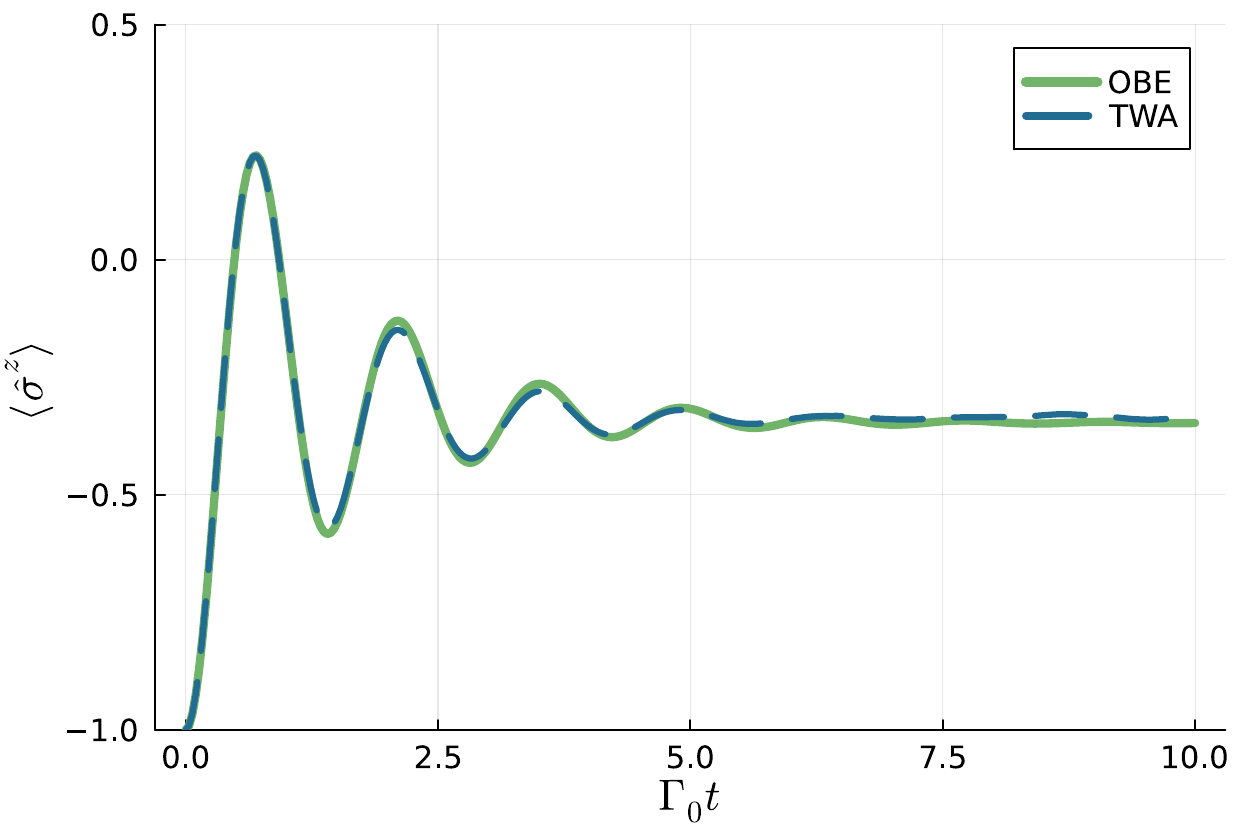}
\end{center}
\vspace{-5mm}
\caption{Expectation value of $\hat \sigma_z$ of a driven single spin starting in the ground state calculated with the optical Bloch equations (green) as well as the \ac{twa} result (blue) for $10^4$ trajectories. Here $\Omega = 2 \Gamma_0$, and $\Delta = 1 \Gamma_0$.}
\label{fig:single Spin Dynamics}
\end{figure}

Let us first discuss how the coherent dynamics of a single spin, as well as decay and dephasing can be described in phase space.
\begin{itemize}
\item[(1)] A \textbf{coherent driving} of a spin with Rabi frequency $\Omega$ and detuning $\Delta$, described by a Hamiltonian
    \begin{equation*}
        H = \frac{\Omega}{2}\hat\sigma^x + \frac{\Delta}{2} \hat \sigma_z
    \end{equation*}
    can be translated to \acp{ode} without any truncation approximation:
    \begin{eqnarray*}
    \text{d} \theta &=& - \Omega \sin{\phi} \, \text{d}t\\
    \text{d} \phi &=& -[ \Omega \cot{\theta} \cos{\phi} \, -\Delta]\text{d}t.
    \end{eqnarray*}
\item[(2)] \textbf{Spontaneous decay / incoherent pump} from the upper to the lower spin state or vice versa, described by a Lindbladian
    \begin{eqnarray*}
        {\cal L}_\downarrow\rho &=& \frac{\Gamma_{0}}{2} \left (2 \hat \sigma^-\rho \hat \sigma^+ -\{\hat \sigma^+\hat \sigma^-,\rho\}\right),\\
        {\cal L}_\uparrow\rho &=& \frac{\Gamma_{0}}{2} \left (2 \hat \sigma^+\rho \hat \sigma^- -\{\hat \sigma^-\hat \sigma^+,\rho\}\right),
    \end{eqnarray*}
   translates into the following \acp{sde}
   \begin{eqnarray*}
    \text{d} \theta &=& \Gamma_0 \Big( \cot{\theta} \pm \frac{1}{\sqrt{3}} \csc{\theta} \Big) \text{d}t,\\
    \text{d} \phi &=& \sqrt{\Gamma_0 \big( 1 + 2 \cot^2{\theta} \, \pm \frac{2}{\sqrt{3}} \csc{\theta} \cot{\theta} \big)} \, \text{d}W.
    \end{eqnarray*}
    Here the upper sign corresponds to decay from spin up to spin down and the lower sign for the opposite process. One notices that the dynamics drives the system 
    to stable steady states with $\cos\theta = \mp 1/\sqrt{3}$ and homogeneous distribution of $\phi$.
\item[(3)] \textbf{Dephasing} of the spin transition governed by
    \begin{equation*}
         {\cal L}_\textrm{deph}\rho = \gamma \left (\hat \sigma^z\rho \hat \sigma^z -\rho\right)
    \end{equation*}
    corresponds to
    \begin{equation*}
        \text{d} \phi = \sqrt{4 \gamma} \, \text{d}W.
    \end{equation*}
\end{itemize}
The dynamics of a single spin, including decay, can be treated exactly in \ac{twa} and deviations between stochastic simulations and exact results are due to sampling noise only.
In  Fig.~\ref{fig:single Spin Dynamics} we have plotted the time evolution of $\langle \hat\sigma_z(t)\rangle$ of a single, coherently driven two-level atom with detuning and spontaneous decay starting in the ground state. One recognizes perfect agreement of the stochastic simulation (apart from some small sampling noise) with exact results, obtained from the two-level Bloch equations, which is the case for any set of parameters.
%

\subsection{Spin-spin interactions}

While the single-spin dynamics including decay and dephasing can be described in phase space exactly without further approximation, this is no longer the case for interacting spins.

\begin{itemize}
    \item[(4)] The equations of motion of the Wigner function for an \textbf{Ising interaction} between spins of the form
    \begin{equation*}
        H = -\sum_{ij} \frac{J_{ij}}{2}\hat \sigma_i^z \hat \sigma_j^z,
    \end{equation*}
    contains higher-order derivatives as well as non-positive contributions in the diffusion matrix and must be truncated. Neglecting higher-order derivatives then leads to the following approximate \acp{ode}
    \begin{eqnarray*}
    \text{d} \theta_i &=& 0, \\
    \text{d} \phi_i &=& - 2 \sqrt{3} \sum_{j \neq i} J_{ij}\cos\theta_j \, \text{d}t.
\end{eqnarray*} 
  \item[(5)] A similar truncation is needed when considering the dynamical equation of the Wigner function for an \textbf{xy-interaction} of the form
    \begin{equation*}
         H= -\sum_{i\neq j} h_{ij} 
    \hat \sigma_i^+ \hat \sigma_j^- +h.a.,
    \end{equation*}
   which eventually yields the deterministic equations
   \begin{eqnarray*}
    \text{d} \theta_i &=& -2 \sqrt{3} \sum_{j\neq i} h_{ij}\sin{\theta_j} \sin{\left( \phi_i -\phi_j \right)},  \\
    \text{d} \phi_i &=& -2 \sqrt{3}\cot\theta_i \sum_{j\neq i} h_{ij}\,\sin{\theta_j} \cos{\left( \phi_i -\phi_j \right)}.
    \end{eqnarray*}
\end{itemize}

\subsection{Dissipative generation of entanglement}

Despite being a semiclassical approximation, the \ac{twa} can describe the generation of entanglement 
due to coupling between spins. We have already seen that the maximally entangled Bell states have a positive
Wigner representation in continuous phase space, thus a dynamical process starting from a factorized initial state
with positive Wigner function generating Bell states could in principle be  described by \acp{sde}. To show that this
is indeed the case, we consider the generation of entanglement between two atoms with Dicke-like collective spontaneous emission.
In the two-atom Dicke model, only symmetric states are coupled to the quantized light field and can decay. Thus dissipation couples
the states
\begin{equation}
\vert \! \uparrow \rangle \vert \! \uparrow\rangle\,\,  \longrightarrow \,\, \vert \Psi_+\rangle \,\, \longrightarrow \,\, \vert \! \downarrow \rangle \vert \! \downarrow\rangle
\end{equation}
with twice the single-atom decay rate, while the Bell state $\vert \Psi_-\rangle$ is completely decoupled, i.e. a dark state. Initially exciting one atom only
\begin{equation}
\vert \psi(t=0)\rangle = \vert \! \uparrow\rangle \vert \! \downarrow\rangle = \frac{1}{\sqrt{2}} \Bigl(\vert \Psi_+\rangle + \vert \Psi_-\rangle\Bigr),
\end{equation}
leads, at large times, to an entangled mixed state with probability $0.5$ being in the Bell state $\vert \Psi_-\rangle$
\begin{equation}
\hat\rho(t\to\infty) = 0.5 \, \vert \! \downarrow\downarrow\rangle \langle \downarrow\downarrow \! \vert + 0.5 \, \vert \Psi_-\rangle\langle \Psi_-\vert.
\end{equation}
This state has a non-vanishing concurrence of $C=0.5$, which is a quantitative measure of entanglement between two spins~\cite{hill1997entanglement}.
It is determined by the square roots of the eigenvalues of $\hat{\rho}\hat{\tilde{\rho}}$, where $\hat{\tilde{\rho}}$ denotes the partial transpose of $\hat\rho$, and $\lambda_1,\dots, \lambda_4$ are the ordered eigenvalues of $\hat{\rho}\hat{\tilde{\rho}}$,
\begin{align}
    \label{eq:Concurrence-Rhotilde}
    \hat{\tilde{\rho}} &= \left(\sigma_y\otimes\sigma_y\right) \hat{\rho}^* \left(\sigma_y\otimes\sigma_y\right),
    \\
    \label{eq:Concurrence-definition}
    C &= \mathrm{max}\left( 0, \sqrt{\lambda_1} - \sqrt{\lambda_2} - \sqrt{\lambda_3} - \sqrt{\lambda_4}\right).
\end{align}

\begin{figure}[h]
\begin{center}
\includegraphics[width=0.48\textwidth]{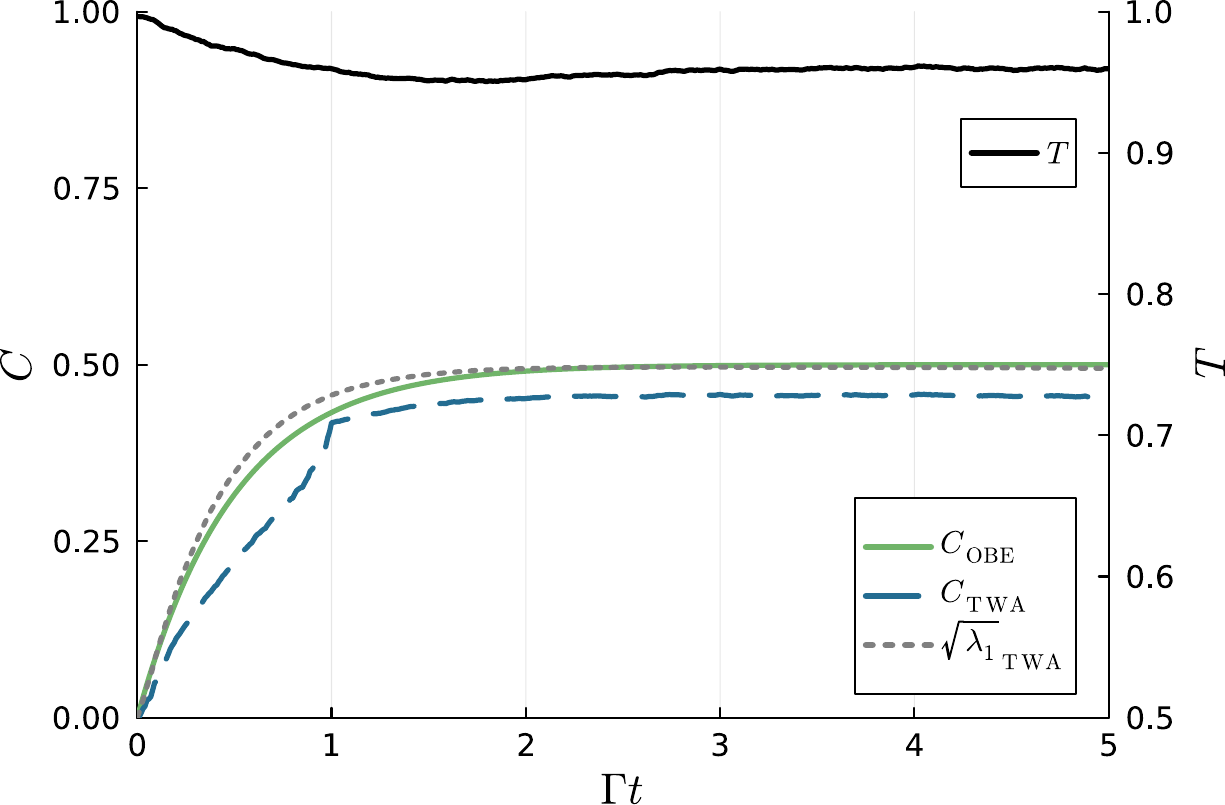}
\end{center}
\vspace{-5mm}
\caption{
Comparison of two-spin concurrence $C$ (left axis) and trace distance $T$ (right axis) between exact dynamics from 2-particle \ac{obe} and \ac{twa} simulation upon Dicke decay. The initial state is $\vert \! \uparrow \rangle \vert \! \downarrow\rangle$. The grey, dotted line corresponds to $\sqrt{\lambda_1}$ from eq.~\eqref{eq:Concurrence-definition} obtained via \ac{twa}.}
\label{fig:Concurrence-Dicke}
\end{figure}

In Fig.~\ref{fig:Concurrence-Dicke} we have plotted the concurrence as function of time obtained from the exact dynamics and
TWA. Here, we applied the collective correspondence rules, as these are to be used in the specific case of Dicke decay. 
Through the approximations made to map the \ac{eom} of the flattened Wigner function to \acp{sde} some of the eigenvalues obtained in TWA become slightly negative for $\Gamma t \ge 1$ and were set to zero, which causes the cusp in the concurrence.
Note that the error of the collective correspondence rules decreases as the number of spins increases. Nevertheless, reasonably good agreement is observed, showing the potential of the continuous \ac{twa} to even describe processes where entanglement is created between spins.

To verify whether the state reproduced by \ac{twa} is indeed close to the exact solution, we additionally examined the trace distance between the exact 
($\hat \rho$) and the approximated density matrix ($\hat \omega$) obtained by \ac{twa}:
\begin{equation}
    \label{eq:Fidelity-Dicke-decay}
    D = \frac{1}{2}\vert\vert \hat \rho -\hat \omega\vert\vert = \frac{1}{2}\tr\Bigl\{ \sqrt{\left( \hat{\rho} - \hat{\omega}\right)^\dagger \left( \hat{\rho} - \hat{\omega} \right)} \Bigr\}.
\end{equation}
In Fig.\ref{fig:Concurrence-Dicke} we have plotted $T=1-D$ as a black curve.
Despite the slightly negative eigenvalues of $\hat \omega$, the  \ac{twa} solution  ends up at a deviation of only approximately $4\%$ from the exact one. Moreover, in the specific case of Dicke decay, $\lambda_1$ should be the only non-vanishing eigenvalue and thus $C=\sqrt{\lambda_1}$. The gray, dotted line shows $\sqrt{\lambda_1}$ from \ac{twa} as function of time, which is smooth and closer to the exact concurrence.
%

\subsection{Collective atom-light interaction}

An important example of collective spin interactions, where the approximate correspondence rules from Sec.~\ref{sect:approximate} can be applied, is superradiance from a collection of atoms~\cite{dicke1954coherence,gross1982,rehler1971superradiance,gross1976observation,skribanowitz1973observation}. If  two-level emitters are localized to a volume with linear dimensions small compared to the wavelength of the resonant transition, their collective interaction with the quantized electromagnetic field can be described by the celebrated Dicke model~\cite{dicke1954coherence}, for which many results can be derived analytically or with little numerical effort. This is no longer the case if the spatial size of the system becomes larger than the wavelength.
Here, inhomogeneous radiative couplings between the atoms need to be taken into account. If retardation effects are neglected, the atom-light interaction can be mapped to an effective spin model by formally integrating out the quantized radiation field, see e.g.
\cite{lehmberg1970radiation,dzsotjan2010quantum,asenjo2017atom,chang2018colloquium}. 
The resulting Lindblad master equation for $N$ such collectively coupled spins driven by an external laser field with Rabi frequency $\Omega$ reads
\begin{align} 
    \frac{\text{d} }{\text{d} t}\hat{\rho} = -i[\hat{H}, \hat{\rho}] +
    \sum_{mn}^N \frac{\Gamma_{mn}}{2} \Bigl(2 \hat{\sigma}_m^- \hat{\rho} \hat{\sigma}_n^+ -\{ \hat{\sigma}_m^+ \hat{\sigma}_n^-, \hat{\rho} \} \Bigr),
    \label{eq:wellenleiter_mastergleichung}
\end{align}
which contains a non-local dissipative coupling $\Gamma_{mn}$ and a Hamiltonian contribution
\begin{equation}
    \hat{H} = -\Omega(\hat{S}^+ + \hat{S}^-) +
    \sum_{mn}^N \frac{J_{mn}}{2} \left(\hat{\sigma}_m^+ \hat{\sigma}_n^- +
        \hat{\sigma}_n^+ \hat{\sigma}_m^-\right),
    \label{eq:wellenleiter_hamiltonian}
\end{equation}
where $\hat S^\mu = \sum_n c_n \, \hat \sigma^\mu_n$.
The radiative couplings $J_{nm}$ and the decay matrix $\Gamma_{mn}$ are given by the Green's tensor $\mathbf{D}$ of the electric field
\begin{equation}
    -J_{mn} + \frac{i}{2} \Gamma_{mn} = \frac{1}{\epsilon_0} \left( \frac{2 \pi \omega_e}{c} \right) \mathbf{p}^\dagger \cdot \mathbf{D}(\boldsymbol{r}_m, \boldsymbol{r}_n, \omega_e) \cdot \mathbf{p},
\end{equation}
 with $\mathbf{p}$ being the dipole vector of the two-level transition, and $\boldsymbol{r}_m, \boldsymbol{r}_n$ the positions of the spins.
Applying the collective correspondence rules~\eqref{coll_weyl_truncated} 
we end up with the following \acp{sde}:
\begin{align}
    \text{d} \theta_n = &\bigg\{ \sqrt{3} \sum_{m=1}^N \sin{\theta_m} \bigg[ \sin{\phi_{mn}} \left( J_{mn}' + \frac{\Gamma_{mn}}{2}'' \right) \notag \\
    &+ \cos{\phi_{mn}} \left( \frac{\Gamma'_{mn}}{2} + J_{mn}'' \right) \bigg] + \frac{\Gamma_{nn}}{2} \cot{\theta_n} \bigg\} \text{d} t \notag\label{eq:theta_collective} \\
    &+ \sum_{m=1}^N \bigg[ \left( - G_{nm}' \cos{\phi_n} + G_{nm}'' \sin{\phi_n} \right) \, \td W_{\theta_m} \notag \\
    &\qquad + \left( G_{nm}' \sin{\phi_n} + G_{nm}'' \cos{\phi_n} \right) \, \td W_{\phi_m} \bigg] ,
\end{align}
\begin{align}  
    \text{d} \phi_n = &\bigg\{ \sqrt{3} \sum_{m=1}^N \sin{\theta_m} \bigg[ \cos{\phi_{mn}} \left( -J_{mn}' + \frac{\Gamma_{mn}}{2}'' \right) \notag \\
    &+ \sin{\phi_{mn}} \left( \frac{\Gamma_{mn}'}{2} - J_{mn}'' \right) \bigg] \bigg\} \cot{\theta_n} \, \td t \notag \label{eq:phi_collective} \\
    &+ \sum_{m=1}^N \cot{\theta_n} \bigg[ \left( G_{nm}' \sin{\phi_n} + G_{nm}'' \cos{\phi_n} \right) \, \td W_{\theta_m} \notag \\
    &\qquad + \left( G_{nm}' \cos{\phi_n} - G_{nm}'' \sin{\phi_n} \right) \, \td W_{\phi_m} \bigg], 
\end{align}
where the coupling matrices
\begin{align}
    \Gamma &= \Gamma' + i \, \Gamma'', \\
    \mathrm{J} &= \mathrm{J}' + i \, \mathrm{J}''
\end{align}
are split up in their real and imaginary parts and $\Gamma$ is decomposed according to $\Gamma = GG^\top$. For  real valued couplings the above equations result in the same \acp{sde} as shown in~\cite{mink2023collective}.

\subsubsection{Dicke superradiance}
\begin{figure}[h!]
\begin{center}
\includegraphics[width=0.48\textwidth]{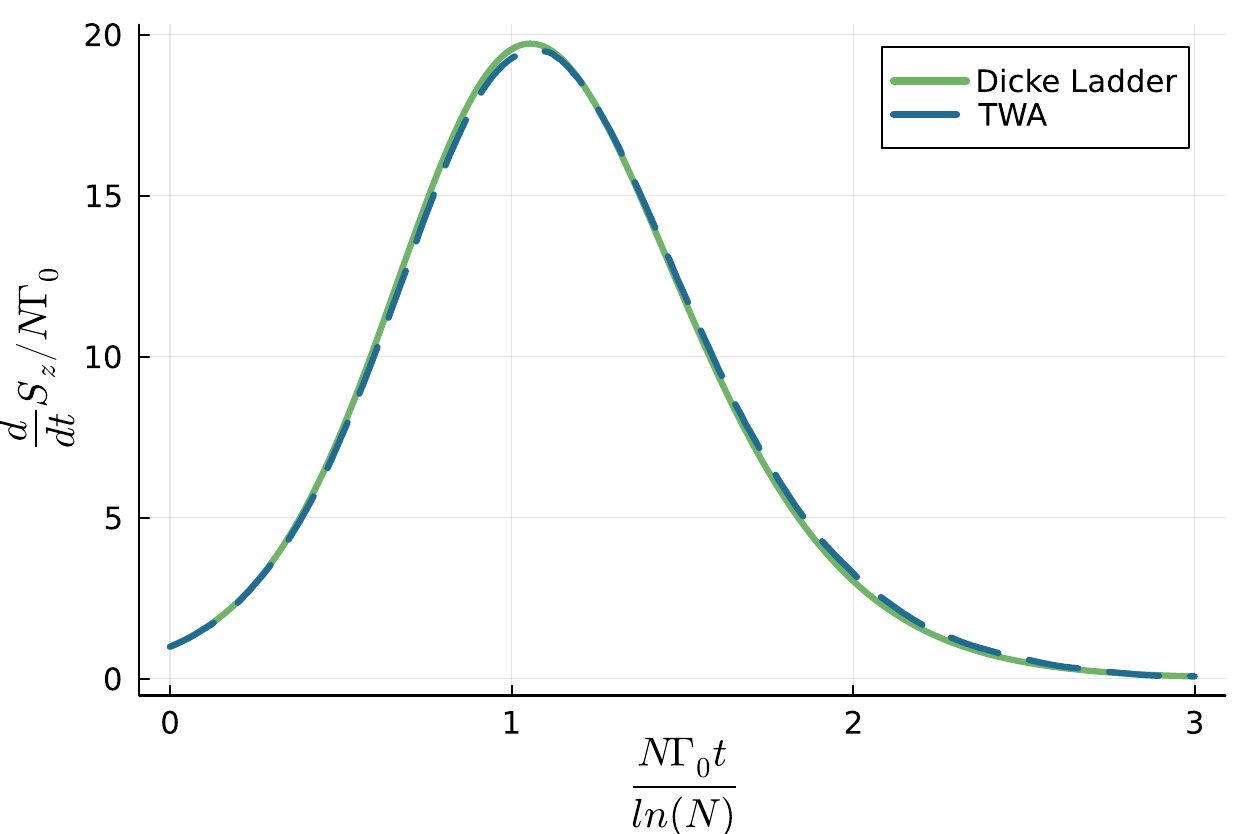}
\end{center}
\vspace{-5mm}
\caption{Emitted intensity of the initially ($t=0$) fully excited Dicke model for 100 atoms. Shown is the \ac{twa} result with $2\cdot10^3$ trajectories compared to the rate equation solution for the Dicke ladder.}
\label{fig:Dicke_Model}
\end{figure}

In Fig.~\ref{fig:Dicke_Model} we plot the emission dynamics of an initially inverted ensemble of 100 two-level atoms 
in a volume small compared to the resonant wavelength, obtained by \ac{twa} along with the results obtained from direct solutions of the Dicke model. In the small-volume limit the unitary contributions are neglected.
One recognizes superradiant behavior and very good agreement between \ac{twa} and exact simulations. As discussed in~\cite{mink2023collective}, it should be noted, however, that while superradiance is perfectly captured by the approximate collective correspondence rules, the \ac{twa} fails to accurately describe sub-radiance.

\subsubsection{Collective non-Gaussian light emission from a driven ensemble of two-level emitters}

While the collective emission of light from an ensemble of atoms in a very small volume can easily be calculated by several methods, the theoretical treatment of spatially extended ensembles is much more involved.
In Refs \cite{tebbenjohannsPRA2024,spahn2026motioninduceddirectionalitycollectiveemission} \ac{twa} simulations have been successfully employed for such problems in waveguides. 
Motivated by the experimental investigation of possible non-Gaussian correlations in the emitted light from an extended, elongated ensemble of two-level atoms driven by an external laser, in Ref.\cite{PhysRevLett.132.133601}, we now explore this question theoretically using the \ac{twa}.
In the experiment, the second-order correlation function 
of the emitted light was detected using a Hanbury-Brown Twiss setup. In the steady state, a reduction of the $g^{(2)}$ correlations below the value of two, expected for Gaussian states, was observed, while the average electric field is zero at the detector. The reduction increases with growing atom number, while it approximately remains constant over a large range of the driving strength.

\begin{figure}[h]
    \centering
        \includegraphics[width=0.48\textwidth]{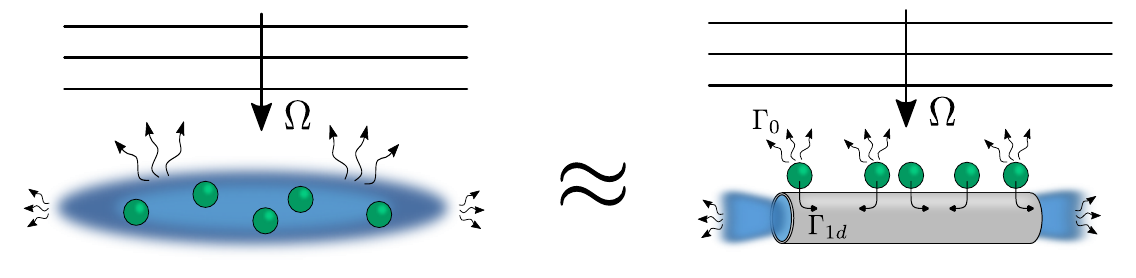}
        \caption{Cigar-shaped ensemble of atoms driven perpendicular to the z-axis (left) and approximate one-dimensional waveguide model used for \ac{twa} simulations (right). }\label{fig:waveguide}
\end{figure}

In order to understand this observation, we simulate an effective  model instead of the three-dimensional problem. The cigar-shaped gas in the experiment has an expansion along the z-axis of 40-50 wavelengths, while the perpendicular diameter is less than a single wavelength. Since we are only interested in the light emitted along the $z$-direction, we use a one-dimensional waveguide model as illustrated in Fig.~\ref{fig:waveguide}.

In the waveguide model, the cooperative emission of the atoms is described by a collective coupling to the 1D continuum of  modes propagating along the waveguide, while spontaneous emission and perpendicular driving are accounted for as single atom processes. Here, the matrices for collective interaction $J_{mn}$ and decay $\Gamma_{mn}$
simplify to 
\begin{align}
    J_{mn} - \frac{i}{2}\Gamma_{mn} =
        -i\frac{\Gamma_\textrm{1D}}{2}\exp{\Bigl( \frac{2\pi i}{\lambda} \lvert z_n - z_m \rvert \Bigr)}.
    \label{eq:wellenleier_green}
\end{align}
$\Gamma_\textrm{1D}$ is an effective coupling strength. 


Assuming that only the two-level transition couples to the propagating quantum field, the electric field, respectively the creation and annihilation operators of photons directly correspond to the spin operator $\hat{\sigma}_n^-$ and $\hat{\sigma}_n^+$~\cite{PhysRevResearch.4.023026}. Therefore we can calculate the electric field by taking the weighted sum over all spins in the system. In the case of our one dimensional model this amounts to
\begin{align}
    \hat{E}^-(t) &= \sum_{n=1}^N c_n \, \hat{\sigma}_n^-(t) \nonumber \\
    &= \sum_{n=1}^N i \sqrt{\Gamma_n} \, e^{\pm i k_0 z_n} \, \hat{\sigma}_n^-(t) = \hat{S}^-(c_n) (t),
\label{eq:wl_el_feld_koll_spin}
\end{align}
where the decay rates $\Gamma_n$ are absorbed into the weights $c_{n}$.

\ 
Now, we can express the observables of the emitted light within our approximation of collective operators by using the truncated correspondence rules~\eqref{coll_weyl_truncated}.
The emitted intensity then yields

\begin{equation}
    \langle \hat{I} \rangle = \langle \hat{S}^+( c_n ) \hat{S}^-( c_n ) \rangle \approx \overline{ \lvert \Sc^-(c_n ) \rvert^2 + \Sc^z( \lvert c_n \rvert^2 )}.
    \label{eq:int_koll}
\end{equation}
%

\begin{figure}[h]
    \centering
        \includegraphics[width=0.48\textwidth]{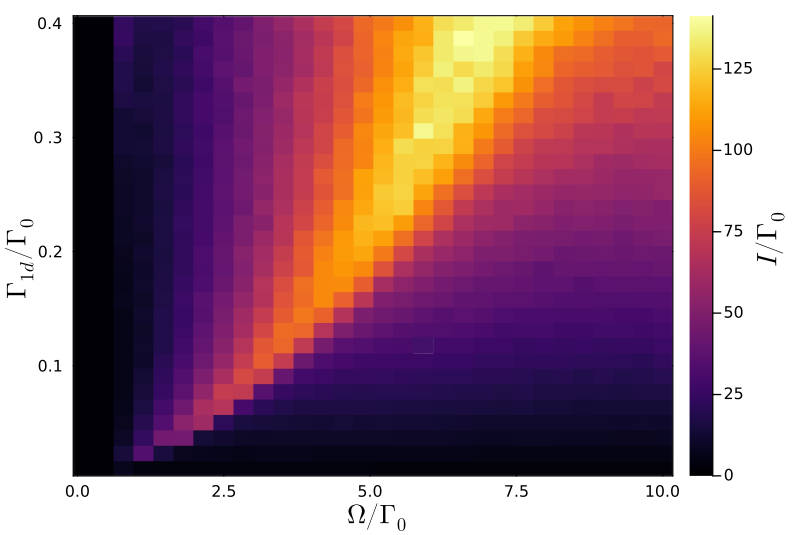}
    \caption{Emitted steady-state intensity from 1000 coupled atoms in the waveguide model plotted over coupling strength and driving. Average over $5\cdot10^3$ trajectories.}
    \label{fig:wl_int_200}
\end{figure}

\begin{figure}[h]
    \centering
        \includegraphics[width=0.48\textwidth]{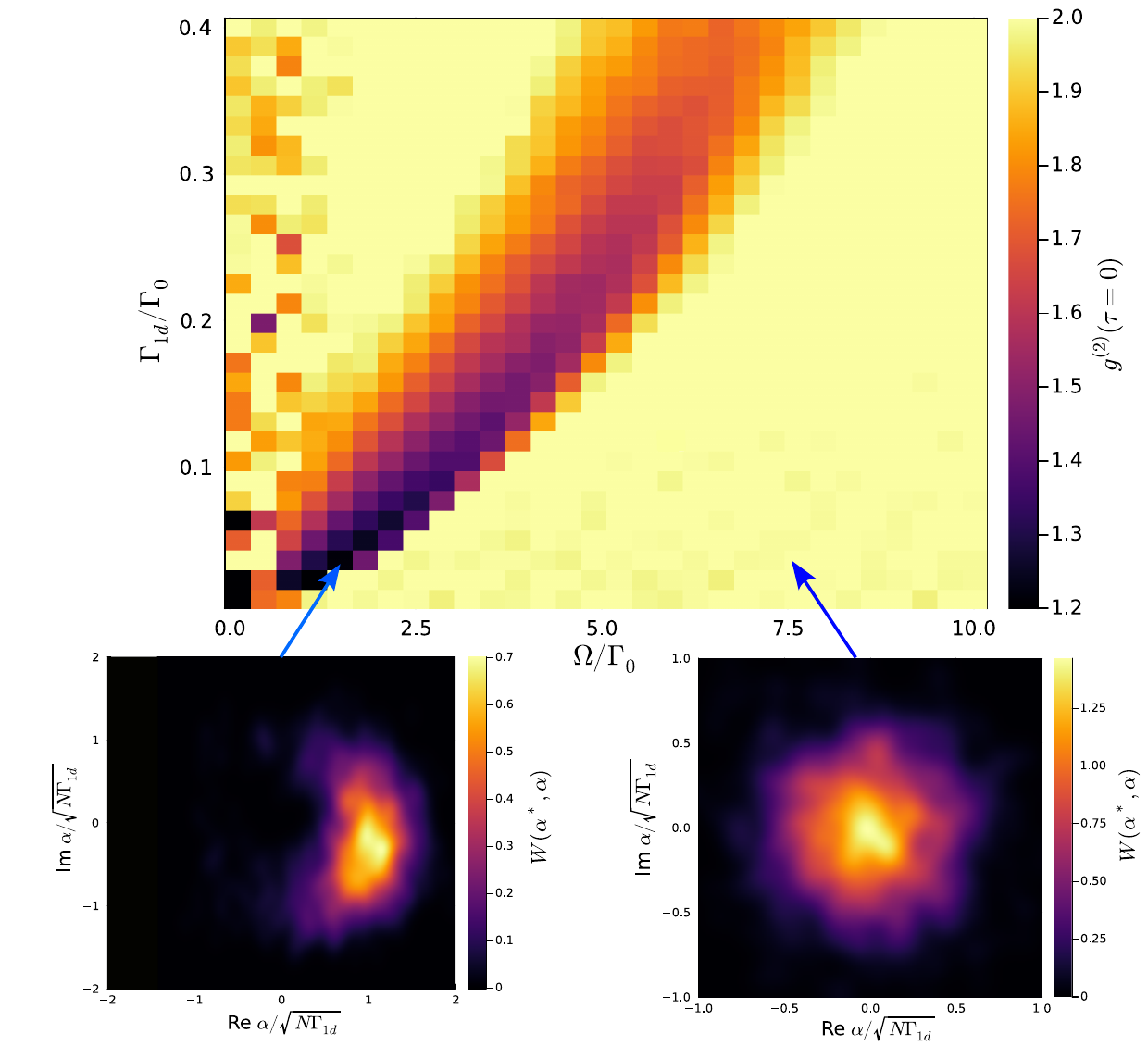}
    \caption{Second order correlation function of the emitted light and Wigner function of the electric field exemplary for two sets of parameters for 1000 atoms.}
    \label{fig:wl_wigner_in_g2}
\end{figure}

The numerator of the normalized second order correlation function $g^{(2)}$ for $\tau = 0$ is a product of four operators and therefore yields
\begin{widetext}
\begin{align}
    &\langle \hat{E}^+ \hat{E}^+ \hat{E}^- \hat{E}^- \rangle \propto \langle \hat{S}^+(c_n ) \hat{S}^+(c_n) \hat{S}^-(c_n) \hat{S}^-(c_n) \rangle  \label{eq:g2_koll_num}\\ \notag \\ &\quad\approx \overline{ \lvert \Sc^-(c_n) \rvert^4 + 4 \Sc^z(\lvert c_n \rvert^2) \, \lvert \Sc^-(c_n) \rvert^2 + 2 (\Sc^z(\lvert c_n \rvert^2))^2 - \text{Re}\{ \Sc^-(c_n) \, 
    \Sc^- (\lvert c_n \rvert^2 c_n^*) \} }. \notag
\end{align}
With these approximations we can calculate the second order correlation function of the light emitted by the atoms as
\begin{align}
    g^{(2)}(\tau = 0) \approx \frac{ \overline{ \lvert \Sc^-( c_n ) \rvert^4 + 4 \Sc^z(\lvert c_n \rvert^2 ) \, \lvert \Sc^-( c_n ) \rvert^2 + 2 (\Sc^z( \lvert c_n \rvert^2 ))^2 - \text{Re}\{ \Sc^-( c_n ) \, 
    \Sc^- ( \lvert c_n \rvert^2 c_n^* ) \}}}{\left( \, \overline{ \lvert \Sc^-(c_n) \rvert^2 + \Sc^z(\lvert c_n \rvert^2) } \, \right)^2}.
    \label{eq:g2_koll}
\end{align}
\end{widetext}
%
%
When expressing the time in units of the spontaneous decay rate $\Gamma_0$, two parameters remain: the driving strength $\Omega$ and the coupling to the waveguide $\Gamma_{1D}$.
The system shows superradiant bursts for short times and subsequently reaches a steady state, whose properties depend on the ratio between those two parameters. 
In Fig.~\ref{fig:wl_int_200} the steady state intensity $\hat{I} = \langle \hat{E^\dagger} \hat{E} \rangle$ is plotted as function of the two parameters.
%
For small driving or small coupling, none to little intensity is emitted into the waveguide mode. The steady state emission increases and later decreases with both parameters.
However, there is an optimal ratio between those parameters, where the intensity takes on the highest values.
To further investigate this, we plot the second order correlation function of the emitted light in Fig.~\ref{fig:wl_wigner_in_g2}.
%

One can clearly distinguish regions of Gaussian light emission  with $g^{(2)} = 2$ outside the regime of maximum emission intensity (see Fig.~\ref{fig:wl_int_200}), in which the correlation function decreases below the value of 2. The Wigner functions of the emitted light clearly show a Gaussian distribution for the uncorrelated case and a crescent moon in the region of non-Gaussian emission. The average phase of the emitted light is determined by the phase of the Rabi drive at the emission time.

\color{black}

\section{Two-time correlations} \label{sec:Two-time}

Since phase space methods map the time evolution of the density operator to that of a quasi-probability distribution in phase space, the calculation of two-time (or multipe-time) correlations is in general difficult. Moreover specific operator orderings required in the mapping between observables and c-number expressions, such as symmetric operator ordering for Wigner functions of bosonic fields, and the time ordering considered in multi-time correlations conflict each other, and require special treatment~\cite{Berg}. Here we show how two-time correlation functions of spin observables can be simulated in an efficient way using the \ac{twa} approach for spins~\cite{Hartmann-diplom}. Note that a related approach was put forward in~\cite{guardiola2025phase}. 

\subsection{Quantum regression theorem}


Two-time correlations can be obtained in the Schr\"odinger picture using the quantum regression theorem, see e.g.~\cite{gardiner2004quantum}. The starting point is their expression
in the Heisenberg picture:
\begin{equation}
    \langle \hat A(t+\tau) \hat B(t)\rangle = \tr\left\{\hat A_H(t+\tau) \hat B_H(t) \hat \omega\right\}
\end{equation}
where $\hat \omega=\textrm{const}_t$ is the time-independent density operator of the total system consisting of spins and reservoir, and $ \hat A_H(t+\tau)  = e^{iH (t+\tau)} \hat A e^{-iH (t+\tau)}$, and $\hat B_H(t)  = e^{iH t} \hat B e^{-iH t}$.
Thus
\begin{equation}
    \langle \hat A(t+\tau) \hat B(t)\rangle = \tr\left\{ e^{iH \tau} \hat A e^{- iH \tau}\hat B \hat \omega(t)\right\},
\end{equation}
where
\begin{align}
    \hat\omega(t)= e^{-iH t} \, \hat \omega\,  e^{iH t}.
\end{align}
Note, that if one chooses $t=0$, i.e. the initial time, calculating the correlation function amounts to simulating a single-time observable in ''state'' $\hat B\hat \omega$, which can be done directly, see e.g. Ref.~\cite{shirai2025out}.
Since $\hat A$ and $\hat B$ are operators acting only in the Hilbert space of spins, we can carry out the trace over the reservoir degrees of freedom
\begin{equation}
    \langle \hat A(t+\tau) \hat B(t)\rangle = \tr_S\left\{  \hat A
    \, \tr_B\left\{e^{- iH \tau}\hat B \hat \omega (t)e^{iH \tau}\right\}\right\}.
\end{equation}
We now define the operator $\hat Y(\tau,t)$ 
\begin{equation}
    \hat Y(\tau,t) = e^{- iH \tau}\hat B \hat \omega(t)e^{iH \tau},\quad \hat Y(0,t)=\hat B\hat \omega(t),
\end{equation}
which corresponds to the total (system $+$ reservoir) density matrix when considering simple expectation values. Tracing over the reservoir degrees of freedom gives an operator in the Hilbert space of the spins, which takes over the role of the reduced
density operator in simple expectation values $ \hat X(\tau,t) = \tr_B\{\hat Y(\tau,t)\} $. It
fulfills the \ac{eom} with respect to the difference time $\tau$:
\begin{equation}
    \frac{d}{d\tau} \hat X(\tau,t) = -i \,\tr_B\bigl\{[H,\hat Y]\bigr\}.
\end{equation}
This equation is the basis of the quantum regression theorem. Now
making  a Born-Markov approximation in the system-reservoir coupling, which amounts to $\hat \omega(t)=\rh(t)\otimes \rh_B(t)$ leads to the Lindblad equation~\eqref{eq:Lindblad} for $\hat X(\tau,t)$ in $\tau$ 
\begin{equation}
    \frac{\partial}{\partial \tau}  \hat X(\tau,t) = {\cal L} \hat X(\tau,t)
\end{equation}
with the solution
\begin{equation}
    \hat X(\tau,t) = e^{{\cal L}\tau} \hat X(0,t) =  e^{{\cal L}\tau} \left[\hat B \rh(t)\right].\label{eq:EOM-X}
\end{equation}
In terms of $\hat X(\tau,t)$ the two time correlation can be expressed as 
\begin{equation}
     \langle \hat A(t+\tau) \hat B(t)\rangle = \tr\left\{  \hat A
    \, \hat X(\tau,t) \right\}.\label{eq:two-time}
\end{equation}

\subsection{Phase-space representation of two-time correlations}

The two-time correlation, eq.~\eqref{eq:two-time}, can be expressed in terms of phase space variables
\begin{align}
     &\langle \hat A(t+\tau) \hat B(t)\rangle  = \tr\left\{  \hat A
    \, \hat X(\tau,t) \right\}\nonumber\\
    &\qquad = \int\!\! d\Om\, {\cal A}(\Om) \tr\left\{  \Dm(\Om)
    \, \hat X(\tau,t) \right\} \\
     &\qquad = \int\!\! d\Omt\,  {\cal A}(\Om) {\cal X}(\Om,\tau,t),\nonumber 
\end{align}
where in the last step we have introduced the flattend Weyl symbol ${\cal X}$ of $\hat X$.

Since the \ac{eom} of $\hat X(\tau,t)$ with respect to $\tau$ is exactly the same as the Lindblad equation of $\rh(t)$, its Weyl symbol ${\cal X}(\Om,\tau,t)$ at relative time $\tau$ obeys
\begin{eqnarray}
    {\cal X}(\Om,\tau,t) = \int \!\! d\Omt^\prime\, p(\Om,\tau; \Om^\prime,0)\, {\cal X}(\Om^\prime,0,t).
\end{eqnarray}

The remaining task is to determine ${\cal X}(\Om^\prime,0,t)$. To this end we note that $\hat X(0,t) = \hat B \rh(t)$
\begin{align}
    & {\cal X}(\Om^\prime,0,t) = \Bigl[\prod_j\pi \sin\theta_j^\prime\Bigr]\, 
    \tr\left\{ \hat B \rh(t) \Dm(\Om^\prime)\right\}
    \\
   &\qquad = \Bigl[\prod_j\pi\sin\theta_j^\prime\Bigr]\!  \int\! d\Omt^{\prime\prime}\,  \chi(\Om^{\prime\prime},t) \, \tr\bigl\{\Dm(\Om^\prime) \hat B \Dm(\Om^{\prime\prime}) \bigr\}.\nonumber
\end{align}

 The function
 \begin{equation}
     f(\hat B,\Om^{\prime},\Om^{\prime\prime}) = \tr\Bigl\{\Dm(\Om^\prime) \hat B \Dm(\Om^{\prime\prime}) \Bigr\}
 \end{equation}
can be explicitly evaluated for every operator $\hat B$. Thus we eventually end up at
the following phase space representation for a two-time correlator
\begin{align}
     &\langle \hat A(t+\tau) \hat B(t)\rangle  = \label{eq:two-time-solution}\\
    &\quad = \int\!\!\!\int\!\!\!\int\!\!\! \int\!\!d\Omt \, d\Omt^\prime d\Omt^{\prime\prime} d\Omt^{\prime\prime\prime}\,
    {\cal A}(\Om) \,  p(\Om,\tau; \Om^\prime,0) \times \nonumber\\
    &\qquad \times \, \Bigl[\prod_j\pi \sin\theta_j^\prime\Bigr]\, f(\hat B,\Om^\prime,\Om^{\prime\prime}) \, p(\Om^{\prime\prime},t; \Om^{\prime\prime\prime},0)\, \chi_0(\Om^{\prime\prime\prime}).\nonumber
\end{align}
This expression can be stochastically simulated
by the following procedure:
\begin{itemize}
    \item[(i)] Solve the Ito \acp{sde}~\eqref{eq:SDE} to obtain $\Omega^{\prime\prime}(t)$ with the initial value $\Omega^{\prime\prime}(t=0)=\Omega^{\prime\prime\prime} $
    randomly chosen from the initial distribution $\chi_0(\Omega^{\prime\prime\prime})$.
    \item[(ii)] Choose an arbitrary value of $\Omega^\prime$ on the unit sphere as new initial value for the second time evolution. 
    \item[(iii)] Solve the Ito \acp{sde}~\eqref{eq:SDE} to obtain $\Omega(t+\tau)$ with the initial values $\Omega(t)=\Omega^\prime$ from (ii).\\
    \item[(iv)] Evaluate the expression \\
    ${\cal A}(\Omega(t)) \Bigl(\prod_j\pi \sin\theta_j^\prime\Bigr)\, f(\hat B,\Omega^\prime,\Omega^{\prime\prime}(t))$.
    \item[(v)] Average over many trajectories according to the initial distribution $\chi_0(\Omega^{\prime\prime\prime})$ and the uniform distribution of $\Omega^\prime$'s, and - if the \acp{sde} contain noise terms - also over these noise realizations.
\end{itemize}
This shows that any two-time correlations can, in principle, be simulated. 

\subsection{Two-time correlations of single-spin operators}

For many applications, the relevant two-time correlations are correlations between single-spin operators, in particular $\hat B \to \hat B_j$, where $j$ denotes a particular spin. In this case, the general simulation procedure simplifies significantly as we can write
\begin{align}
    & f(\hat B_j,\Om^\prime,\Om^{\prime\prime}) = \nonumber \\
    & \quad =\tr_j\{\hat \Delta_j(\Omega^\prime) \hat B_j \hat \Delta_j(\Omega^{\prime\prime})\} \prod_{k\ne j} \tr_k\{ \hat \Delta_k(\Omega^\prime)\hat\Delta_k(\Omega^{\prime\prime})\}. \nonumber \\
    & \quad =\tr_j\{\hat \Delta_j(\Omega^\prime) \hat B_j \hat \Delta_j(\Omega^{\prime\prime})\} \\
    & \qquad \times
    \prod_{k\ne j} \Bigl[\delta(\Omega^\prime_k-\Omega_k^{\prime\prime}) - 2\pi \sum_{l>1,m}Y_{l}^{m*}(\Omega_k^\prime) Y_{l}^{m}(\Omega_k^{\prime\prime})\Bigr]. \nonumber
\end{align}
The contributions from the higher spherical harmonics, however, vanish in the exact time evolution, which can be seen as follows: From eq.~\eqref{eq:two-time-solution} one recognizes that $f(\hat B_j,\Om^\prime,\Om^{\prime\prime})$ is integrated over the Wigner function at time $t$, i.e. there appears a term in eq.~\eqref{eq:two-time-solution}
\begin{align}
    &\int\! d\Omt^{\prime\prime} \!\!\int\! d\Omt^{\prime\prime\prime}f(\hat B_j,\Om^\prime,\Om^{\prime\prime}) p(\Om^{\prime\prime},t;\Om^{\prime\prime\prime},0) \chi_0(\Om^{\prime\prime\prime})= \nonumber\\
    & = \int\! d\Omt^{\prime\prime} f(\hat B,\Om^\prime,\Om^{\prime\prime}) \chi(\Om^{\prime\prime},t)=\int\! d\Om^{\prime\prime} f(\hat B,\Om^\prime,\Om^{\prime\prime}) \W(\Om^{\prime\prime},t) \nonumber\\
    & = \int\! d\Om^{\prime\prime} f(\hat B_j,\Om^\prime,\Om^{\prime\prime}) \tr\{\rh(t) \Dm(\Om^{\prime\prime})\}.\nonumber
\end{align}
The last term contains, however, only spherical harmonics $\Y_l^m(\Omega^{\prime\prime})$ with $l=0,1$.
 Thus all terms in $f(\hat B_j,\Om^\prime,\Om^{\prime\prime})$ proportional to higher-order spherical harmonics in $\Om^{\prime\prime}$ vanish after integration $d\Om^{\prime\prime}$. Thus we are allowed to set
\begin{align}
    & f(\hat B_j,\Om^\prime,\Om^{\prime\prime}) = \\
    & \quad =\tr_j\{\hat \Delta_j(\Omega^\prime) \hat B_j \hat \Delta_j(\Omega^{\prime\prime})\} \prod_{k\ne j} \delta(\Omega_k-\Omega_k^\prime). \nonumber
\end{align}

This represents a significant improvement, as only for one spin $"j"$ new trajectories need to be simulated.  In eq.~\eqref{eq:two-time-solution} we have to sample only one new initial value $\Omega_j^\prime$
at time~$t$ from the unit sphere and propagate it with the \acp{sde} to time~$t+\tau$. The trajectories for all other spins are just continued from $t$ to $t+\tau$. In essence, the spin $"j"$ is doubled. 
Its first copy "A" starts with the initial state of the system and is evolved until time $t$. Then its coupling to the other spins is switched off.
The second copy "B" starts in a fully mixed state and remains decoupled until "t"
at which time its coupling to the other spins is switched on.
To calculate two-time correlations of collective variables, e.g. the spectrum of the light emitted by an ensemble of spins, we may need two-time correlations for \emph{all} spins, i.e. $j=1,2,\dots N$. In this case the effective number of spins just doubles.

In terms of stochastic variables, the explicit expression for the evaluation of two-time correlations of single-spin operators $\hat A_k$ and $\hat B_j$ reads
\begin{align}
    \label{eq:2z_mittelwert_final}
    &\langle \hat{A}_k(t+\tau) \hat{B}_j(t) \rangle =  \\ 
    &\quad = \overline{\pi \sin{\Tilde{\theta}_j(t)} \, \mathcal{A}({\Omega}_k(t+\tau)) \, f(\hat B_j, \Tilde{\Omega}_j,\Omega_j(t))}.\nonumber 
\end{align}
Here $\Omega_l(t)=\{\theta_l(t),\phi_l(t)\}$ are the phase-space variables time-evolved from initial values at $t=0$, $\Omega_l(t=0)=\{\theta_{l0},\phi_{l0}\}$,
which are distributed according to the Wigner distribution of the initial quantum state. At time $t$ the phase-space variables $\{\theta_j(t),\phi_j(t)\}$ of the $j$'th spin are reset to random values on the sphere 
$\{\Tilde{\theta}_j,\Tilde{\phi}_j\}$ and the time evolution is continued until time $t+\tau$ with the same \acp{sde}.

\begin{figure}[h]
\begin{center}
\includegraphics[width=0.48\textwidth]{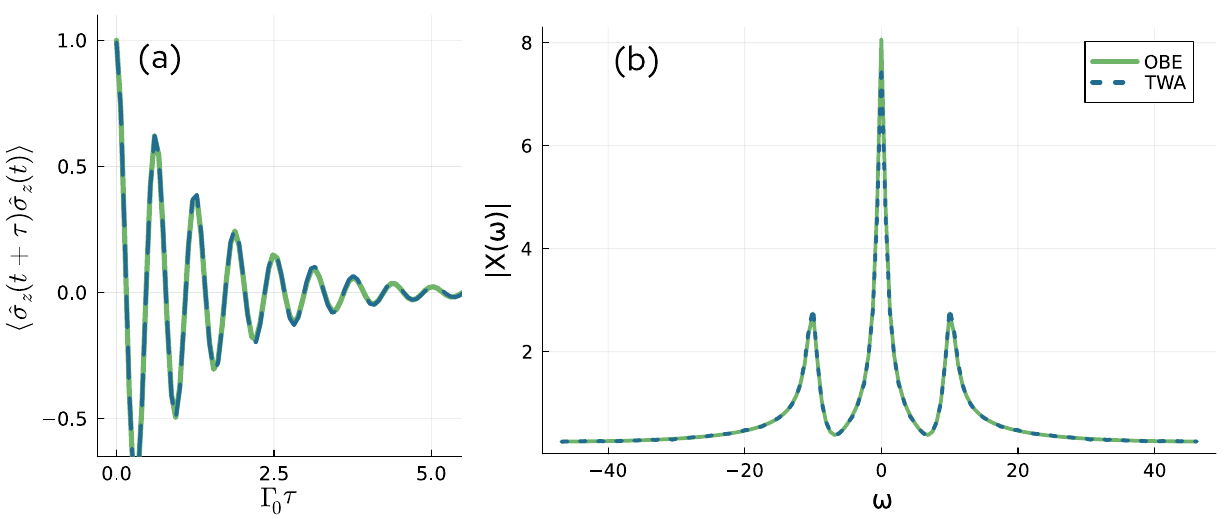}
\end{center}
\vspace{-5mm}
\caption{(a) Two-time $\sigma_z$ correlation function~\eqref{eq:2z_sigma_z_explicit} of a single atom for $\Omega = 5 \, \Gamma_0$, $\Delta = 0$ and $10^5$ trajectories. (b) Mollow triplett for same parameters.}
\label{fig:two-time-single}
\end{figure}

To benchmark the method, we first consider the simple problem 
of a single two-level atom, continuously driven by a resonant laser field with Rabi frequency $\Omega$ and subject to spontaneous decay with rate $\Gamma_0$ from the excited to the ground state. 
In Fig.~\ref{fig:two-time-single}(a) we plot the \ac{twa} result for the two-time correlation function of $z$ components
\begin{align}
    \label{eq:2z_sigma_z_explicit}
    &\langle \hat{\sigma}^z (t + \tau) \hat{\sigma}^z (t) \rangle = \\ \nonumber \\
    &= \overline{\pi \sin{\Tilde{\theta}(t)} \sqrt{3} \cos{(\Tilde{\theta}(t + \tau))} \, \frac{\sqrt{3}}{2} \bigg( \cos{(\Tilde{\theta}(t))} + \cos{(\theta(t))} \bigg)},\nonumber
\end{align}
along with the correlation function obtained from the full density matrix and one recognizes perfect agreement. Additionally, we show the Mollow triplett in Fig.~\ref{fig:two-time-single}(b) which is computed as the Fourier transform of the  two-time correlation $\langle \sigma^+(t+\tau) \sigma^- (t)\rangle$ in the steady state.
Again, perfect agreement with the \ac{obe} is obtained.

Next we consider a system of two two-level atoms coupled via the electric field, with $m,n\in\{1,2\}$:
\begin{align}
    \frac{J_{mn}}{\Gamma_0} &= - \frac{3}{4} \Bigg\{ \left[ 1 - \lvert \hat{e}_p \cdot \hat{e}_{mn} \rvert^2 \right] \frac{\cos{(k_0 r_{mn})}}{k_0 r_{mn}}  \label{eq:3d_j} \\
    & - \left[1 - 3 \lvert \hat{e}_p \cdot \hat{e}_{mn} \rvert^2 \right] \bigg[ \frac{\sin{(k_0 r_{mn})}}{(k_0 r_{mn})^2} + \frac{\cos{(k_0 r_{mn})}}{(k_0 r_{mn})^3} \bigg] \Bigg\}, \nonumber\\
    \nonumber \\
    \frac{\Gamma_{mn}}{\Gamma_0} &= \frac{3}{2} \Bigg\{ \left[ 1 - \lvert \hat{e}_p \cdot \hat{e}_{mn} \rvert^2 \right] \frac{\sin{(k_0 r_{mn})}}{k_0 r_{mn}} \label{eq:3d_gamma}\\
    & + \left[1 - 3 \lvert \hat{e}_p \cdot \hat{e}_{mn} \rvert^2 \right] \bigg[ \frac{\cos{(k_0 r_{mn})}}{(k_0 r_{mn})^2} - \frac{\sin{(k_0 r_{mn})}}{(k_0 r_{mn})^3} \bigg] \Bigg\}, \nonumber
\end{align}
with the distance vector $r_{mn}$ and the unit vector $\hat{e}_p$ of the lights polarization. 

\begin{figure}[h]
\begin{center}
\includegraphics[width=0.48\textwidth]{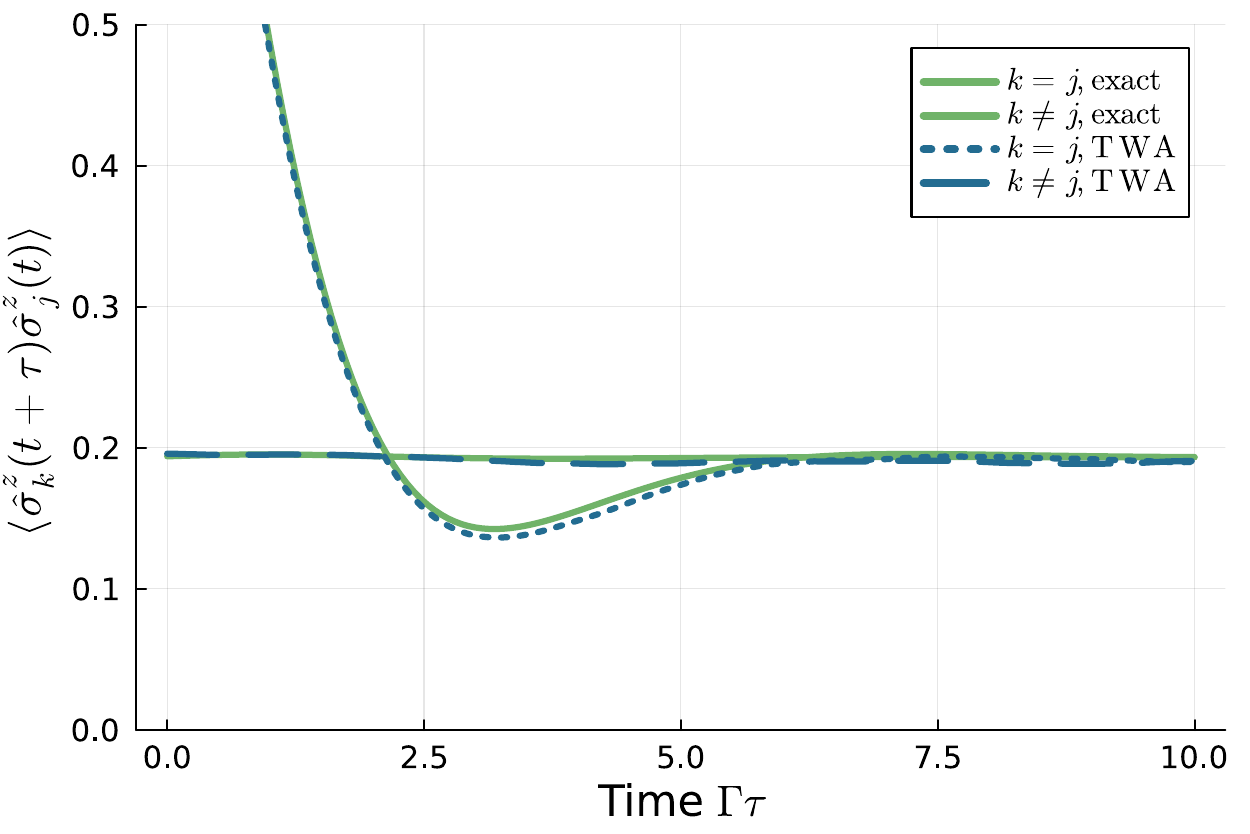}
\end{center}
\vspace{-5mm}
\caption{Two-time correlation function~\eqref{eq:2z_mittelwert_final} of two interacting spins at distance $1.5 \lambda$ for both cases $i = j$ and $i \neq j$ and $\Omega = 0.4 \, \Gamma_0$ averaged over $10^5$ trajectories.}
\label{fig:two-time-two}
\end{figure}

Fig.~\ref{fig:two-time-two} shows the $zz$-correlations for the same spin and for different spins obtained from \ac{twa} and the exact results.
Again, there is close to perfect agreement, showing that two-time correlations between different spins can also be accurately calculated.

\section{Imaginary-time TWA}\label{sec:iTWA}

\subsection{General method}

The \ac{twa} for spins can also be applied to calculate thermal and ground states of interacting spin systems. To this end, it was extended to imaginary time in~\cite{schlegel2026imaginarytimeevolutioninteractingspin}. In the following, we will summarize this \ac{itwa}.

The goal of the \ac{itwa} is to calculate 
 the canonical density matrix at inverse temperature~$\tau=1/k_\text{B}T$
\begin{equation}
    \label{eq:CanonicalDensity}
    \hat \rho(\tau) = \frac{1}{Z(\tau)} e^{-\tau \hat H},
\end{equation}
where $\hat{H}$ is the Hamiltonian and $Z(\tau) = \textrm{Tr}\{e^{-\tau \hat H}\}$ is the partition function.
States at finite temperature or approximations of the ground state can be obtained by evolution  in imaginary time
\begin{equation}
    \label{eq:iTWA-EOM}
    \partial_\tau \hat \rho = -\bigl(\hat H - \langle \hat H\rangle\bigr) \hat \rho = -\frac{1}{2}\bigl\{\hat H,\hat \rho\bigr\}_+ + \langle \hat H\rangle \hat \rho,
\end{equation}
where, due to symmetry reasons, the anti-commutator appears. Applying the mapping between Hilbert and phase space, one finds that the evolution equation of the Wigner function different from eq.~\eqref{eq:W-EOM} now also contains a term proportional to $\chi(\Om,\tau)$ itself:
\begin{align}
    \label{eq:iTWA-PDE}
    \partial_\tau \chi(\Om,\tau) = &-\left[{\cal H}(\Om)-\langle\hat H\rangle(\tau)\right]\chi(\Om,\tau) \nonumber\\
    &-
    \sum_j\frac{\partial}{\partial x_j} A_j(\Om) \chi(\Om,\tau) \\ 
    &+\,\frac{1}{2}\sum_{jl} \frac{\partial^2}{\partial x_j\partial x_l} D_{jl}(\Om) \chi(\Om,\tau) +\cdots\, .\nonumber
\end{align}
Here ${\cal H}(\Om)$ is the Weyl symbol of the Hamiltonian, and $x_k\in\{\theta_j,\phi_j\}$. Truncation at second order derivatives, and approximating the coefficient matrix by a positive definite one, i.e. $\mathbf{D} \to  \mathbf{B}\cdot \mathbf{B}^\top$ leads to a proper Fokker-Planck equation with an additional term $-({\cal H}-\langle\hat H\rangle)\chi$.  As shown in~\cite{schlegel2026imaginarytimeevolutioninteractingspin} expectation values in the canonical state at inverse temperature~$\tau$ can then be calculated by solving a set of stochastic differential equations similar to eq.~\eqref{eq:SDE}
\begin{equation}
    dx_j(\tau) = A_{j} (\Om(\tau))\, d\tau + \sum_l B_{jl}(\Om(\tau))\,  dW_l(\tau).
    \label{eq:iTWA-SDE}
\end{equation}
and stochastic averaging over the trajectories following these \acp{sde}. Due to the term linear in $\chi$ in the \ac{eom} of the flattened Wigner function, thermal averages have, however, a more complicated expression in phase space:
\begin{equation}
    \langle \hat A\rangle(\tau) = \frac{\displaystyle{\overline{{\cal A}(\Om(\tau)) e^{-\int_0^\tau \!\! d\tau^\prime {\cal  H}(\Om(\tau^\prime))} }}}{\displaystyle{\overline{e^{-\int_0^{\tau} \!\! d\tau^{\prime} {\cal  H}(\Om(\tau^{\prime}))} }}}.
    \label{eq:iTWA-expectation-values}
\end{equation}
Now, for each trajectory the Weyl symbol~${\cal A}$ must be weighted with an exponential before the stochastic average is taken. The denominator in eq.~\eqref{eq:iTWA-expectation-values} results from $\langle \hat H\rangle(\tau)$ and corresponds to the partition function of the thermal state. The above expression has some similarity to the expression for expectation values in the canonical ensemble, i.e. $\langle\hat A\rangle(\tau) = \tr\{ \hat A\,e^{-\tau \hat H} \}/Z(\tau)$.
In order to prevent large fluctuations or numerical instabilities, it is useful to replace ${\cal H}(\tau) \to {\cal H}(\tau) - \overline{{\cal H}(\tau)}$ in eq.~\eqref{eq:iTWA-expectation-values}.
%

\subsection{Ising Hamiltonians}

To illustrate the applicability of the \ac{itwa} method, we now determine the finite-temperature state of an (anti-ferromagnetic, AF) Ising Hamiltonian with uniform coupling strength $J$ on a random 3-regular graph with $N=22$ nodes (spins). 
Finding the ground state of such Ising Hamiltonians on general graphs is an important problem in statistical mechanics as it 
is closely related to spin glass physics~\cite{edwards1975theory,parisi1979infinite,binder1986spin} and to the solution of 
combinatorial optimization problems~\cite{kirkpatrick1983optimization,du1998handbook} 
such as MaxCut. This task is in general NP hard (non-deterministic polynomial-time hard), and includes some NP-complete problems~\cite{lucas2014ising}. 
Thus, a general numerical method to calculate ground states in an efficient way is believed to not exist as this would imply NP=P. 

In a 3-regular graph every node (spin) is connected by exactly three edges to other nodes (spins). The corresponding Hamiltonian reads
\begin{equation}
    \label{eq:random-Ising}
    \hat H =  \frac{J}{2}\sum_{i,j\in {\cal G} } \hat\sigma_i^z\, \hat \sigma_j^z,\qquad J>0
\end{equation}
where the sum extends over all spins $(i,j)$ that are connected by edges of the graph ${\cal G}$.

Using $\partial_\phi^2 \hat \Delta$ in the correspondence rules, one recognizes that the $\phi$-dynamics makes no contribution to expectation values of operators whose Weyl symbols are independent of $\phi$, 
\begin{equation*}
    \label{eq:iTWA-tau-derivative}
    \frac{\mathrm{d}\langle\hat A\rangle}{\mathrm{d}\tau} = 
    -\overline{({\cal H}-\langle\hat H\rangle){\cal A}} + \sum_i \overline{A_{\theta_i}\frac{\partial{\cal A}}{\partial \theta_i}}
    + \frac{1}{2} \sum_{i,j} \overline{D_{\theta_i\theta_j} \frac{\partial^2{\cal A}}{\partial\theta_i\partial\theta_j}},
\end{equation*}
and the following drift vector and diffusion matrix are the only relevant terms,
\begin{align*}
    A_{\theta_i} &= - J\left(\frac{2}{\sin\theta_i}-3\sin\theta_i\right) \sum_{i,j\in{\cal G}} \cos\theta_j,  \\
     D_{\theta_i\theta_j} &= -\frac{J}{3}\left(\frac{2}{\sin\theta_i}-3\sin\theta_i\right)\left(\frac{2}{\sin\theta_j}-3\sin\theta_j\right).
\end{align*}
The diffusion matrix is in general not positive and thus must be truncated. Nevertheless, this approximation becomes exact in the limit $\tau\to\infty$ because the deterministic imaginary-time evolution leads to the asymptotics $\sin^2\theta_j \to 2/3$, corresponding to the eigenstates of $\hat\sigma_z$.
At this limit, the diffusion matrix vanishes $\vert D_{\theta_i\theta_j}\vert \to 0$ and the \ac{itwa} becomes exact!
Note, that the latter argument applies to all Ising models with arbitrary couplings $J_{ij}$ and the method can thus be applied to find the ground state of any Ising Hamiltonian.

\begin{figure}[h]
\begin{center}
\includegraphics[width=0.45\textwidth]{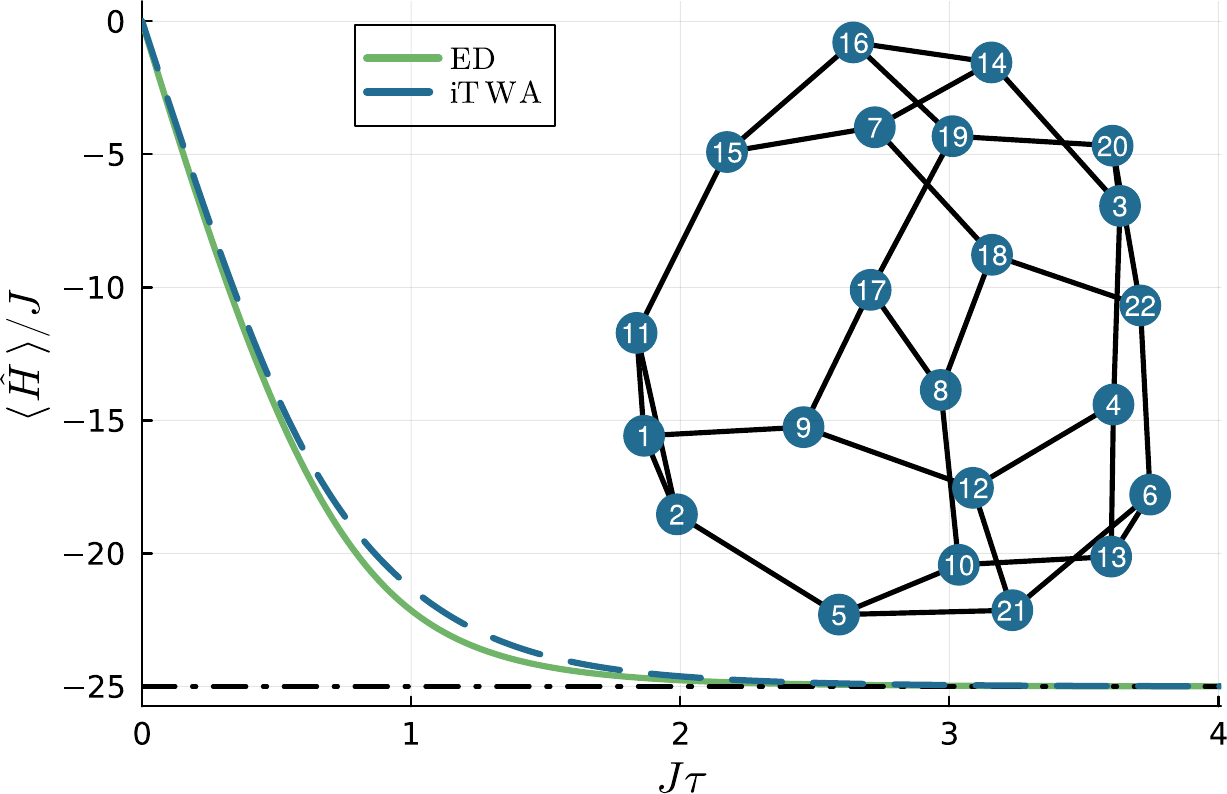}
\end{center}
\vspace{-5mm}
\caption{Average energy $\langle \hat H\rangle$ of the Gibbs state of the AF Ising Hamiltonian, eq.~\eqref{eq:random-Ising}, on a randomly chosen 3-regular graph (see inset) with $N=22$ nodes over $10^5$ trajectories as function of inverse temperature $\tau = 1/k_B T$. Shown are \ac{itwa} simulations (dashed, blue) and results from exact diagonalization (ED, solid, green) as well as the exact ground state energy $E_0=-25J$ (dot-dashed, black).}
\label{fig:iTWA-3ru}
\end{figure}

In Fig.~\ref{fig:iTWA-3ru} we have plotted the average energy~$\langle \hat H\rangle$ of the Gibbs state, eq.~\eqref{eq:CanonicalDensity}, for a randomly chosen 3-regular graph with $N=22$ nodes. Shown are the \ac{itwa} results and values obtained by exact diagonalization (ED). One recognizes very good agreement over the whole range of inverse temperatures. 
While for small imaginary-times the truncation effect is small but noticeable, in the limit $J\tau \geq 3$ the approximation becomes exact and one recognizes a perfect agreement of the ground state energies.

As shown in~\cite{schlegel2026imaginarytimeevolutioninteractingspin}, even for large system sizes $N=\mathcal{O}(100)$ the method achieves very good agreement between the ground state approximated via \ac{itwa} and the exact one subject only to finite-sampling errors. 
If additionally suitable initial distributions are chosen for the fully-mixed state, the diffusion matrix can be kept small $\vert D_{\theta_i\theta_j}\vert < 1$ also for short imaginary-times, and the approximation improves.
It should be noted that the NP-hardness of finding the ground state, reflects itself in an exponential increase of the required number of trajectories with system size~\cite{schlegel2026imaginarytimeevolutioninteractingspin}.

\subsection{Quantum phase transitions}
%
Since the \ac{itwa} can be used to simulate low-temperature states of spin Hamiltonians, the question arises, if it can describe quantum phase transitions (QPT). This has been studied in 
\cite{schlegel2026imaginarytimeevolutioninteractingspin} for the paradigmatic model of the transverse-field Ising model (TFIM):
\begin{equation}
    \label{eq:Transversse-field-Ising}
    \hat H = - h\sum_j\hat \sigma_j^x - \frac{J}{2} \sum_{\langle ij\rangle } \hat\sigma_i^z\, \hat \sigma_j^z.
\end{equation}
It displays a quantum phase transition from a paramagnetic to a ferromagnetic phase at a critical value of $h/J$, depending on the spatial dimension $d$, with critical exponents corresponding to the classical Ising model in $d+1$ dimensions~\cite{sachdev1999quantum}. 

\begin{figure}[h]
\begin{center}
\includegraphics[width=0.48\textwidth]{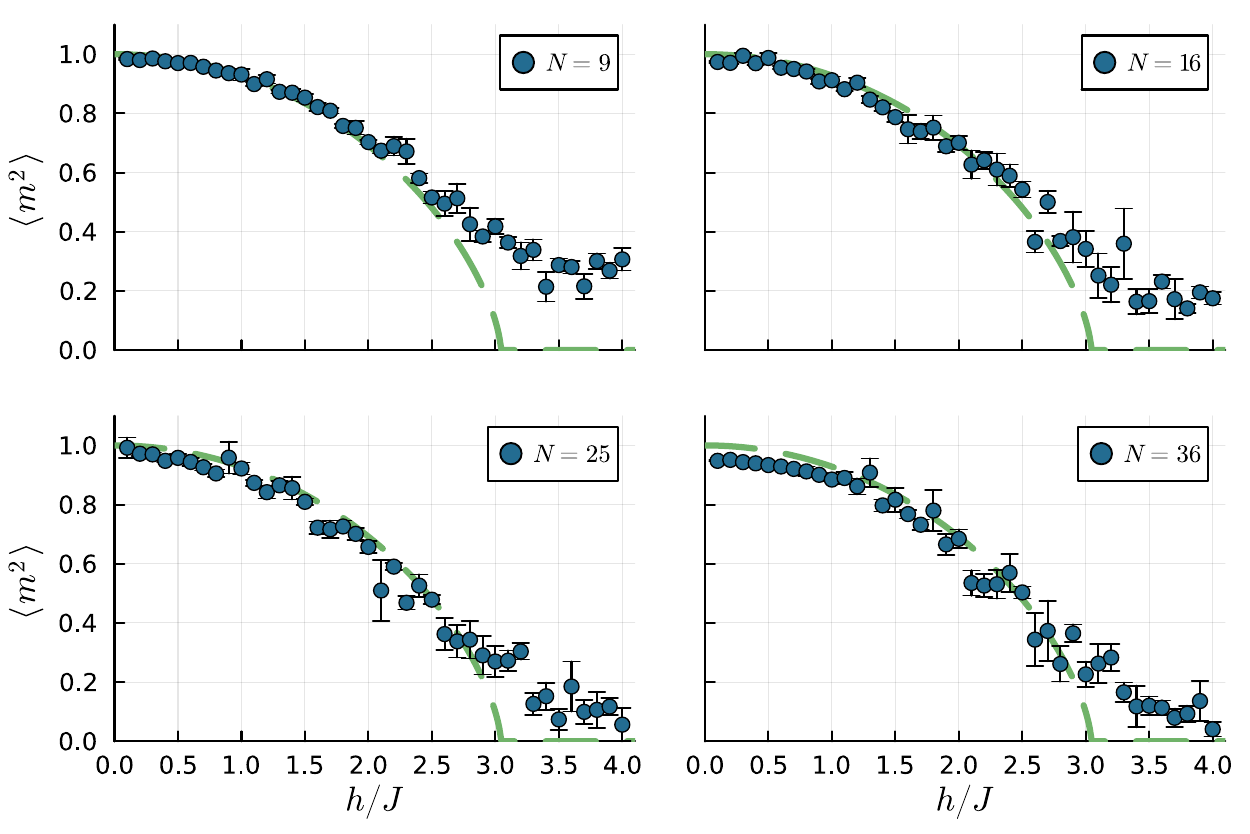}
\end{center}
\vspace{-5mm}
\caption{Order parameter $\langle m^2\rangle$ for the 2D TFIM for different number of spins with exact infinite-size solutions (green dashed line). The stationary values were determined by averaging $\langle m^2\rangle$ over $\tau$ in the interval $J\tau = 1-3$ and $6\cdot 10^5$ trajectories. For the error bars the leave-one-out jackknife standard deviation~\cite{dca15e5b-b3f7-3417-8555-955fe36eb045} was used to account for bias in single trajectories.}
\label{fig:QPT}
\end{figure}

To account for quantum noise in the simulations, we exploit the freedom in the choice of correspondence rules ($\partial_\theta^2$). See~\cite{schlegel2026imaginarytimeevolutioninteractingspin} for details.
In Fig.~\ref{fig:QPT} we have plotted the order parameter $\langle m^2\rangle = (N + \sum_{j\neq i} \langle \sigma_i^z \sigma_j^z \rangle)/N^2$ obtained from \ac{itwa} for sufficiently large values of $J\tau$ on a quadratic 2D lattice with nearest-neighbor interaction, and compared it to exact results in the thermodynamic limit.
The \ac{itwa} describes the phases away from the critical point rather well, near the critical point the simulations show, however, sizable numerical fluctuations for increasing system sizes. Thus, while reproducing the correct qualitative behavior, the \ac{itwa} is only of limited use for studying the quantum critical regime.
%

\section{Path integral derivation}\label{sec:Keldysh}

\begin{figure}[h]
\begin{center}
\includegraphics[width=0.48\textwidth]{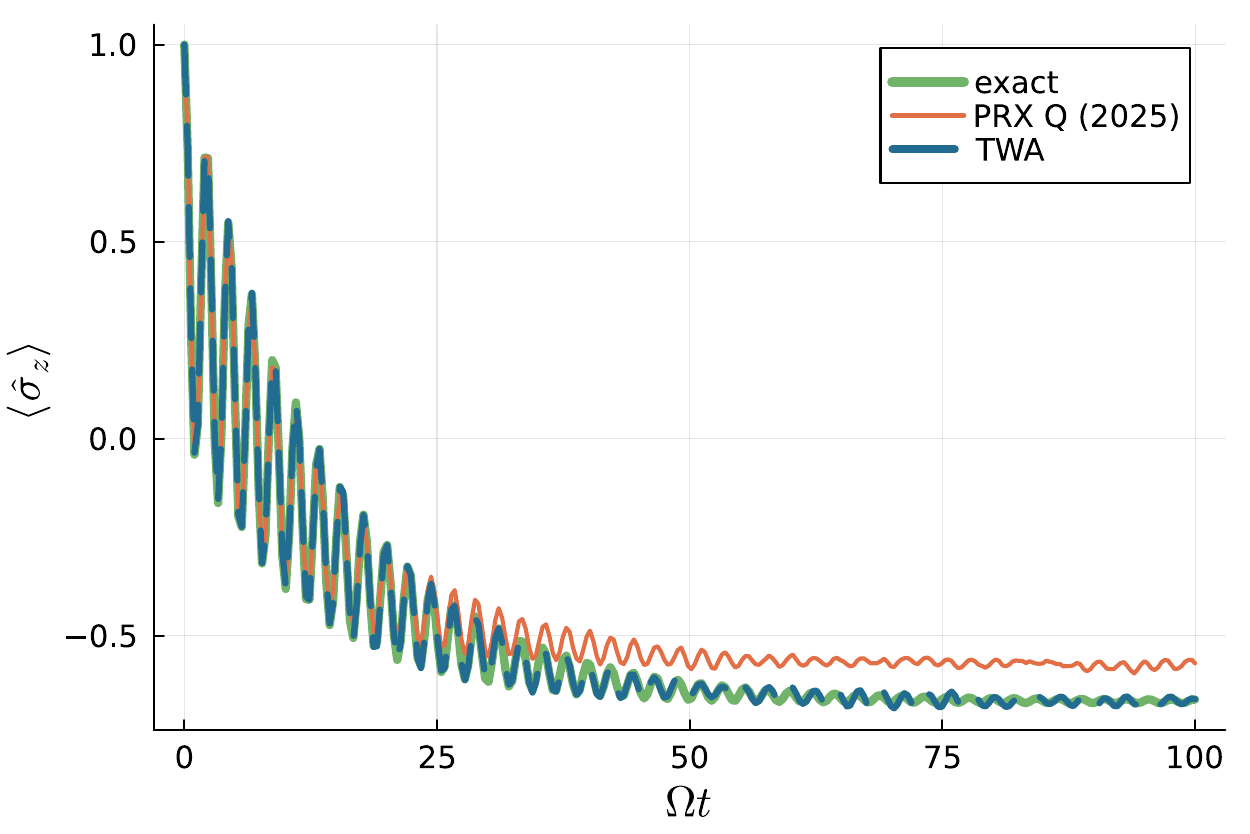}
\end{center}
\vspace{-5mm}
\caption{Population dynamics of a single two-level atom, coherently driven with Rabi frequency $\Omega=10 \Gamma_0$ and detuning $\Delta = 10 \Gamma_0$, with $\Gamma_0$ being the spontaneous decay rate. Shown are the exact solution of the \ac{obe} (green), \ac{twa} simulations with $10^4$ trajectories (blue) and simulations using the \ac{eom} derived in~\cite{hosseinabadi2025user} (orange).}
\label{fig:TWA-single-spin-comparison}
\end{figure}

The equations of motion of the previous sections were obtained by mapping the Lindblad generator onto the Wigner function through a set of correspondence rules. 
There is, however, an equivalent route to the same stochastic dynamics, which recasts the problem as a path integral over phase-space trajectories. 
For open bosonic systems this connection is well established: the \ac{twa} emerges as the truncation of the Keldysh action at quadratic order in the quantum fields, which corresponds to retaining the leading-order term of the semiclassical $\hbar$ expansion~\cite{sieberer2016keldysh, yoneya2025path}.
The corresponding path-integral approach for spin systems has been put forward recently in Ref.\cite{hosseinabadi2025user}.
The resulting stochastic equations for dissipative processes were, however, different from those obtained by mapping the Lindblad generator to phase space and led to small, but non-negligible deviations from exact results. An example is shown in Fig.~\ref{fig:TWA-single-spin-comparison}, where we have compared the \ac{twa} predictions for the single-spin dynamics, which agree with the exact behavior, but differ from the simulation results using the equations of Ref.\cite{hosseinabadi2025user}.
We believe that these discrepancies originate from the subtleties in the mapping of spin operators onto the curved phase space, which modify the treatment of the dissipative terms.
Taking these into account via path integrals, one indeed recovers the stochastic equations derived in Sec.~\ref{sec:SDEs}.
In the following section we outline the main steps of this derivation and details are presented elsewhere~\cite{Noel:2026jra}.

\subsection{The construction of the path integral}

To build the path integral for a single spin, we discretize the Lindblad evolution in time and represent the state in $\mathrm{SU}(2)$ phase space at each step.
To this end, the formal solution of the Lindblad master equation~\eqref{eq:Lindblad} is sliced into $N$ time steps of width $\delta t$
\begin{equation}
    \hat \rho (t) = e^{(t-t_0)\mathcal L}\hat \rho(t_0) = \lim _{N \rightarrow \infty}\left(\mathbb{1}+\delta t \mathcal{L}\right)^N \hat \rho\left(t_0\right),
\end{equation}
so that after the $n$th step, the density matrix becomes
\begin{equation}
    \hat \rho_{n+1} = (1+\delta t \mathcal{L}) \hat \rho_n +\mathcal{O}(\delta t^2).
    \label{eq:next-step-trotter}
\end{equation}
To turn this recursion into a path integral we represent the state at each slice in the phase-space representation of Sec.~\ref{sec:Phase-space}. 
There, the phase-point operator $\hat\Delta(\Omega)$ attached to each point $\Omega=(\theta,\phi)$ of the enlarged Bloch sphere, eq.~\eqref{eq:Delta}, assigns to every operator its $c$-number symbol $O(\Omega)=\tr\{\hat O\,\hat\Delta(\Omega)\}$, eq.~\eqref{eq:Weyl}.
The density matrix is reconstructed from its symbol, the Wigner function $\W(\Omega)=\tr\{\hat\rho\,\hat\Delta(\Omega)\}$, as $\hat\rho=\int d\Omega\,\W(\Omega)\,\hat\Delta(\Omega)$, eqs.~\eqref{eq:A},~\eqref{eq:Wigner}.
Applying this resolution at slice $n$, we get $\hat\rho_n=\int d\theta_n\,d\phi_n\;\W_n(\Omega_n)\,\hat\Delta(\Omega_n)$, where $\W_n$ denotes the weight with respect to the flat measure $d\theta\,d\phi$ \footnote{See~\cite{MinkPRR2022} for details on curved/flat measures.}.
Evolving the system amounts to finding the step from $\W_n$ to $\W_{n+1}$ and iterating over all slices builds the path integral.
Substituting $\hat \rho_n$ into~\eqref{eq:next-step-trotter}, we get
\begin{equation}
\hat\rho_{n+1}=\int d\theta_n\,d\phi_n\;\W_n\,(1+\delta t\,\mathcal L)\,\hat\Delta(\Omega_n).
\label{eq:flat-step}
\end{equation}
The whole construction thus rests on evaluating  $\mathcal L\hat\Delta(\Omega_n)$, for which Sec.~\ref{sec:FPE} has already assembled the necessary ingredients. 
Each term of the Lindbladian multiplies its argument by fixed operators, from the left or from the right. 
Away from the poles, the four operators of eq.~\eqref{eq:set} form a complete basis of the single-spin operator space, so for any operator $\hat X$ there exist unique differential operators $\mathcal D^{L}_{X}$ and $\mathcal D^{R}_{X}$, of finite order in $(\theta, \phi)$ with $c$-number coefficients, such that 
\begin{equation}
\hat X\hat\Delta(\Omega)=\mathcal D^{L}_{X}\hat\Delta(\Omega),
\quad
\hat\Delta(\Omega)\hat X=\mathcal D^{R}_{X}\hat\Delta(\Omega).
\label{eq:onesided}
\end{equation}
These one-sided identities are precisely the direct correspondence rules of Tables~\ref{tab:table-1} and~\ref{tab:table-2}.
They are also the building blocks of the so-called star product: the symbol of an operator product on $\mathrm{SU}(2)$ phase space is not the product of the symbols, but $(\hat O\hat P)_W(\Omega)=(O\star P)(\Omega)\equiv\tr\{\hat\Delta(\Omega)\hat O\hat P\}$, where the ordinary product $O(\Omega) P(\Omega)$ is corrected by a finite number of derivative terms~\cite{varilly1989moyal}.
Applying~\eqref{eq:onesided} term by term to the Lindbladian, we get 
\begin{equation}
\begin{aligned}
\mathcal L\hat\Delta(\Omega_n)
={}&\Bigl[-i\bigl(\mathcal D^{L}_{H}-\mathcal D^{R}_{H}\bigr)
+\sum_\alpha\Bigl(\mathcal D^{L}_{L_\alpha}\mathcal D^{R}_{L^\dagger_\alpha}\\
&-\tfrac12\,\mathcal D^{L}_{L^\dagger_\alpha L_\alpha}
-\tfrac12\,\mathcal D^{R}_{L^\dagger_\alpha L_\alpha}\Bigr)\Bigr]\hat\Delta(\Omega_n).
\end{aligned}
\label{eq:kernel-lindblad}
\end{equation}
%
To chain the slices into a path integral, the update~\eqref{eq:flat-step} must be rewritten as a resolution at the new point $\Omega_{n+1}$,
\begin{equation}
\hat\Delta(\Omega_n)=\int d\theta_{n+1}\,d\phi_{n+1}\,\delta(\Omega_{n+1}-\Omega_n)\,\hat\Delta(\Omega_{n+1}).
\label{eq:deltares}
\end{equation}
Representing the $\delta$-function as a Fourier integral, the update step for $\W_{n+1}$ becomes
\begin{equation}
\W_{n+1}=\!\int\!d\theta_n\,d\phi_n\!\int\!\frac{d\tilde\theta\,d\tilde\phi}{(2\pi)^2}\,
e^{\,i[\tilde\theta\Delta\theta+\tilde\phi\Delta\phi]}
\bigl[1+\delta t\,\ell\bigr]\,\W_n,
\label{eq:stepweight}
\end{equation}
which introduces the pair $(\tilde\theta,\tilde\phi)$ conjugate to the elementary step, and where $\Delta\theta\equiv\theta_{n+1}-\theta_n$, $\Delta\phi\equiv\phi_{n+1}-\phi_n$. 
The generator symbol $\ell$ is read off termwise from~\eqref{eq:kernel-lindblad},
\begin{equation}
\begin{aligned}
\ell(\Omega;\tilde\theta,\tilde\phi)
={}&-i\bigl(H_+-H_-\bigr)
+\sum_\alpha\Bigl[
\bigl(L_\alpha\!\cdot\!L^\dagger_\alpha\bigr)_{+-}\\
&-\tfrac12\bigl(L^\dagger_\alpha L_\alpha\bigr)_+
-\tfrac12\bigl(L^\dagger_\alpha L_\alpha\bigr)_-\Bigr],
\end{aligned}
\label{eq:ell}
\end{equation}
where in each one-sided symbol the derivatives are replaced by their conjugate variables, $\partial_\theta\to-i\tilde\theta$, $\partial_\phi\to-i\tilde\phi$, so that $\ell$ is a polynomial in $(\tilde\theta,\tilde\phi)$.
In~\eqref{eq:ell}, $X_\pm(\Omega;\tilde\theta,\tilde\phi)$ denote the polynomials obtained from $\mathcal D^{L}_{X}$ and $\mathcal D^{R}_{X}$ under this substitution, and $(L_\alpha\!\cdot\!L^\dagger_\alpha)_{+-}$ is the polynomial representing the first term of the dissipative contribution to the Lindbladian~\eqref{eq:Lindblad}.
Chaining the $N$ slices using $1+\delta t\,\ell=e^{\delta t\,\ell}+\mathcal O(\delta t^2)$, the weight of a path collects into a single exponential $e^{iS}$,
\begin{equation}
\W_N=\!\int\!\prod_{n=0}^{N-1}\frac{d\theta_n\,d\phi_n\,d\tilde\theta_n\,d\tilde\phi_n}{(2\pi)^2}\;
e^{\,iS}\;\W_0,
\label{eq:pathint}
\end{equation}
with the action
\begin{equation}
S=\int\!dt\,\bigl(\tilde\theta\,\dot\theta+\tilde\phi\,\dot\phi\bigr)
-i\!\int\!dt\;\ell(\Omega;\tilde\theta,\tilde\phi)
\;\equiv\;S_0+S_H+S_{L}.
\label{eq:Stotal}
\end{equation}
Expectation values follow from eq.~\eqref{eq:expectation} as $\langle\hat O\rangle=\int d\theta_N\,d\phi_N\,\W_N\,O(\Omega_N)$.
In~\eqref{eq:Stotal}, $S_0$ collects the Fourier phases of~\eqref{eq:stepweight}, while the remainder is split into the Hamiltonian $S_H$ and the dissipative $S_L$ contribution.

\subsection{Connection to the semiclassical limit of Keldysh field theory}

The action~\eqref{eq:Stotal} carries twice as many variables per time slice as the sphere has coordinates, namely the propagated configuration $(\theta,\phi)$ and the ``response pair'' $(\tilde\theta,\tilde\phi)$ introduced with the Fourier representation~\eqref{eq:stepweight}.
A doubling of this kind is characteristic of two closely related field theoretic frameworks.
On the one hand, it is the structure of the so-called MSRJD (Martin-Siggia-Rose-Janssen-Dominicis) path integral of stochastic dynamics~\cite{martin1973statistical, janssen1976lagrangean, de1976technics}, where an auxiliary response field is introduced for every physical variable to linearize the evolution equation, and integrating over it enforces the stochastic \ac{eom}.
On the other hand, in Keldysh field theory, every degree of freedom appears once on the forward and once on the backward branch of the closed time contour, as the density matrix is evolved from both sides~\cite{sieberer2016keldysh}.
A Keldysh rotation recasts the two branches as a classical and a quantum field, and truncating the action at quadratic order in the latter reduces the Keldysh action to MSRJD form, with the quantum fields playing the role of response fields~\cite{sieberer2016keldysh}.
The analogous two-sided structure is in the sidedness of the operators in~\eqref{eq:kernel-lindblad}, i.e.\ in the $\pm$ labels of~\eqref{eq:ell}.

\subsubsection{Coherent contributions}

The commutator of any spin-1/2 Hamiltonian with the phase-point operator contains only first derivatives of $\hat\Delta$.
Assembling the entries of Table~\ref{tab:table-1}, it reads
\begin{equation}
[\hat H,\hat\Delta]
=\frac{2i}{\sqrt3\,\sin\theta}
\bigl(\partial_\phi H\,\partial_\theta\hat\Delta
-\partial_\theta H\,\partial_\phi\hat\Delta\bigr),
\label{eq:spinbracket}
\end{equation}
so that under $\partial_\theta\to-i\tilde\theta$, $\partial_\phi\to-i\tilde\phi$, we get
\begin{equation}
H_+-H_-=-\frac{2}{\sqrt3\,\sin\theta}
\bigl(\tilde\phi\,\partial_\theta H-\tilde\theta\,\partial_\phi H\bigr).
\label{eq:sided-split}
\end{equation}
The sided difference is thus exactly linear in the response fields.
We therefore change variables: the propagated configuration becomes the ``classical'' field, $\theta_{\rm cl}=\theta$, $\phi_{\rm cl}=\phi$, and the response pair is traded for the ``quantum'' fields,
\begin{equation}
\theta_q=-\frac{2\tilde\phi}{\sqrt3\,\sin\theta},\quad
\phi_q=\frac{2\tilde\theta}{\sqrt3\,\sin\theta},
\label{eq:clq-def}
\end{equation}
under which $H_+-H_-=\theta_q\,\partial_\theta H+\phi_q\,\partial_\phi H$, so that $S_H$ is 
\begin{equation}
S_H
=-\int\! dt\,\bigl(\theta_q\,\partial_\theta H_{\rm cl}
+\phi_q\,\partial_\phi H_{\rm cl}\bigr),
\label{eq:SH-clq}
\end{equation}
which is exact for a single spin-$1/2$.
The names are justified by the branch reading of this result: under the standard Keldysh rotation, the classical field is the mean of the two branch copies and the quantum field their difference, $\theta_\pm=\theta_{\rm cl}\pm\theta_q/2$ and $\phi_\pm=\phi_{\rm cl}\pm\phi_q/2$.
Taylor-expanding the branch difference about the classical configuration, the even orders cancel and the leading term reproduces the sided difference above: left (right) multiplication is evaluated on the forward (backward) branch, exactly as on the Keldysh contour.
For the coherently driven spin of Sec.~\ref{sec:SDEs}, the Weyl symbols of the Pauli operators follow from eq.\eqref{eq:CartesianSpin}, giving
$H(\theta,\phi)=\sqrt3\,\bigl[\tfrac{\Omega}{2}\sin\theta\cos\phi+\tfrac{\Delta}{2}\cos\theta\bigr]$, whose derivatives entering~\eqref{eq:SH-clq} are
\begin{equation}
\begin{aligned}
\partial_\theta H_{\rm cl}
&=\sqrt3\,\Bigl[\tfrac{\Omega}{2}\cos\theta_{\rm cl}\cos\phi_{\rm cl}
-\tfrac{\Delta}{2}\sin\theta_{\rm cl}\Bigr],\\
\partial_\phi H_{\rm cl}
&=-\tfrac{\sqrt3\,\Omega}{2}\sin\theta_{\rm cl}\sin\phi_{\rm cl}.
\end{aligned}
\end{equation}

The kinetic term of~\eqref{eq:Stotal} follows by direct substitution of~\eqref{eq:clq-def},
\begin{equation}
S_0=\int\!dt\,\tfrac{\sqrt3}{2}\sin\theta_{\rm cl}
\bigl(\phi_q\,\dot\theta_{\rm cl}-\theta_q\,\dot\phi_{\rm cl}\bigr).
\label{eq:S0-clq}
\end{equation}
%

\subsubsection{Dissipative contributions}

We now evaluate the dissipative action for the spontaneous-decay channel of Sec.~\ref{sec:SDEs}, $\hat L=\sqrt{\Gamma_0}\,\hat\sigma^-$, whose symbol follows from Table~\ref{tab:table-1}: $L(\theta,\phi)=(\sqrt{3\Gamma_0}/2)\sin\theta\,e^{-i\phi}$.
Expanded in the derivative basis~\eqref{eq:set} via Table~\ref{tab:table-2}, each dissipator entry is a multiple of $\hat\Delta$ itself plus derivative terms. 
This multiple is the same for $\hat\sigma^-\hat\Delta\hat\sigma^+$ and $\tfrac12\{\hat\sigma^+\hat\sigma^-,\hat\Delta\}$ (both equal the symbol of
$\hat\sigma^+\hat\sigma^-$, i.e.\ the star product
$(\sigma^+\!\star\sigma^-)(\Omega)=\tfrac12(1+\sqrt3\cos\theta)$) so it cancels out.
The dissipator applied to $\hat\Delta$ then consists of derivative terms only,
\begin{equation}
\Gamma_0\Bigl[\hat\sigma^-\hat\Delta\,\hat\sigma^+
-\tfrac12\{\hat\sigma^+\hat\sigma^-,\hat\Delta\}\Bigr]
=A_\theta\,\partial_\theta\hat\Delta
+\tfrac12 B_{\phi\phi}\,\partial_\phi^2\hat\Delta,
\label{eq:diss-Delta}
\end{equation}
with
\begin{equation}
\begin{aligned}
A_\theta&=\Gamma_0\Bigl(\cot\theta+\tfrac{\csc\theta}{\sqrt3}\Bigr),\\
B_{\phi\phi}&=\Gamma_0\Bigl(1+2\cot^2\theta
+\tfrac{2}{\sqrt3}\csc\theta\cot\theta\Bigr),
\end{aligned}
\label{eq:diss-coeffs}
\end{equation}
i.e.~of pure drift and diffusion, with precisely the coefficients of the decay equations quoted in Sec.~\ref{sec:SDEs}.
The substitution $\partial_\theta\to-i\tilde\theta$, $\partial_\phi\to-i\tilde\phi$ converts~\eqref{eq:diss-Delta} into the dissipative part of the step weight, $\ell_{\rm diss}=-i A_\theta\,\tilde\theta-\tfrac12 B_{\phi\phi}\,\tilde\phi^{\,2}$, which is exact as the basis~\eqref{eq:set} carries at most two derivatives.
Rewriting in the classical and quantum fields, the dissipative action becomes
\begin{equation}
\begin{aligned}
S_{L}^{\rm cl/q}
={}&-\!\int\! dt\,\phi_q\,\tfrac{\Gamma_0}{2}\bigl(1+\sqrt3\cos\theta_{\rm cl}\bigr)\\
&+\tfrac{3i\Gamma_0}{8}\!\int\! dt\,\theta_q^{2}
\Bigl[1+\cos^2\theta_{\rm cl}+\tfrac{2}{\sqrt3}\cos\theta_{\rm cl}\Bigr].
\end{aligned}
\label{eq:Sdiss-clq}
\end{equation}
The real term, linear in $\phi_q$, is the dissipative drift; the imaginary term, quadratic in $\theta_q$, encodes the noise. 
Had the star products instead been replaced by ordinary products of the branch symbols, like in the truncation route of the bosonic constructions~\cite{sieberer2016keldysh} applied to spins in~\cite{hosseinabadi2025user}, both the drift and the noise would acquire spurious corrections \footnote{By contrast, the dephasing channel of Sec.~\ref{sec:SDEs} is insensitive to this: its jump operator is unitary and Hermitian, so the exact dissipator is a rigid $\pi$-rotation of the phase-point operator (pure azimuthal diffusion, with no drift for star-product corrections to shift) and the ordinary-product route yields the same diffusion. The equations of motion acquire a white noise of variance $4\gamma$ for $\dot\phi_{\rm cl}$\cite{hosseinabadi2025user, noel2026quantum}.}. 

\subsubsection{Equations of motion}

Collecting the coherent~\eqref{eq:SH-clq},~\eqref{eq:S0-clq} and the dissipative parts~\eqref{eq:Sdiss-clq}, the full action reads
\begin{equation}
S=\int\!dt\,\Bigl[\theta_q\,\mathcal E_\theta
+\phi_q\,\mathcal E_\phi
+\tfrac{i}{2}\,D\,\theta_q^{2}\Bigr],
\label{eq:S-canonical}
\end{equation}
with
\begin{equation}
\begin{aligned}
\mathcal E_\theta&=-\tfrac{\sqrt3}{2}\sin\theta_{\rm cl}\,\dot\phi_{\rm cl}
-\partial_\theta H_{\rm cl},\\
\mathcal E_\phi&=\tfrac{\sqrt3}{2}\sin\theta_{\rm cl}\,\dot\theta_{\rm cl}
-\partial_\phi H_{\rm cl}
-\tfrac{\Gamma_0}{2}\bigl(1+\sqrt3\cos\theta_{\rm cl}\bigr),\\
D&=\tfrac{3\Gamma_0}{4}\Bigl[1+\cos^2\theta_{\rm cl}
+\tfrac{2}{\sqrt3}\cos\theta_{\rm cl}\Bigr].
\end{aligned}
\label{eq:EED}
\end{equation}
Since only the $\theta_q^{2}$ entry of the noise kernel is nonzero, the quadratic term is decoupled by a single real auxiliary field $\xi(t)$ with Gaussian weight $\exp[-\tfrac12\!\int\!dt\,\xi^{2}/D]$ ~\cite{martin1973statistical,janssen1976lagrangean}, i.e.\ white noise with correlator $\langle\xi(t)\xi(t')\rangle=D(\theta_{\rm cl})\,\delta(t-t')$, under which the action becomes linear in both quantum fields, $S=\int\!dt\,[\theta_q(\mathcal E_\theta-\xi)+\phi_q\,\mathcal E_\phi]$.
Integrating over $\theta_q$ and $\phi_q$ then produces $\delta$-functionals enforcing $\mathcal E_\theta=\xi$ and $\mathcal E_\phi=0$: the partition function reduces to an average over trajectories obeying these two equations.
Solving them for the time derivatives gives the stochastic equations of motion of the driven-decaying spin,
\begin{equation}
\begin{aligned}
\dot\theta_{\rm cl}&=-\Omega\sin\phi_{\rm cl}
+\Gamma_0\Bigl(\cot\theta_{\rm cl}+\tfrac{\csc\theta_{\rm cl}}{\sqrt3}\Bigr),\\
\dot\phi_{\rm cl}&=\Delta-\Omega\cot\theta_{\rm cl}\cos\phi_{\rm cl}
+\frac{2}{\sqrt3\sin\theta_{\rm cl}}\,\xi(t),
\end{aligned}
\label{eq:eom-keldysh}
\end{equation}
which coincide, drift and noise alike, with the single-spin \acp{sde} for coherent driving and spontaneous decay quoted in Sec.~\ref{sec:SDEs}.

\section{Summary and outlook}

The present paper provided a comprehensive overview of the truncated Wigner approximation for spins, a semiclassical method
that allows one to simulate the dynamics and thermodynamics of 
large interacting, dissipative spin systems efficiently beyond the mean-field level~\cite{MinkPRR2022,mink2023collective}. 
The method is based on a mapping between the many-particle Hilbert space and an SU(2) continuous phase space, where operators are replaced by c-number functions, called Weyl symbols, and the quantum state is represented by a real-valued quasi-probability distribution -- the Wigner function. 

In contrast to approaches based on discrete representations of spins~\cite{DTWA}, the use of a continuous phase-space
provides a way to systematically derive dynamical equations for the Wigner function and to include dissipative processes. 
Under controlled conditions these can be approximated by a \ac{fpe}, which allows to simulate the quantum dynamics in a numerically inexpensive way in terms of \acp{sde}. The number of coupled equations scales only linearly with system size and in many cases the number of trajectories needed to reach a desired precision increases only as a low-degree polynomial.
Thus the \ac{twa} is numerically much less costly than, e.g., cumulant expansions
\cite{kramer2015generalized,Kerber2025}.

Furthermore, the representation of the density matrix in continuous phase space has a gauge degree of freedom, which can be exploited to find positive definite Wigner functions for a much larger class of quantum states, including entangled many-particle states. This is important, since a Fokker-Planck dynamics can map positive definite initial quasi-probability distributions only onto positive definite final ones~\cite{Risken}. 

The \acp{sde} are obtained through a set of correspondence rules, which are not unique and can be adjusted to the problem at hand. The direct correspondence rules, derived in~\cite{MinkPRR2022}, apply to most spin problems. They can also be derived within a path integral approach~\cite{hosseinabadi2025user}, if in the treatment of dissipative processes~\cite{sieberer2016keldysh}
the proper mapping of spin operators onto the curved phase space is used, see~\cite{Noel:2026jra}.
For collective spin processes, such as superradiant emission of light, a different, approximate 
set of collective correspondence rules is often more adequate and easier to apply~\cite{mink2023collective,tebbenjohannsPRA2024,spahn2026motioninduceddirectionalitycollectiveemission}.

Using a phase-space analog of the quantum regression theorem, it is furthermore possible to calculate multi-time correlation functions with small numerical overhead~\cite{Hartmann-diplom,guardiola2025phase}. This allows one to simulate spectra and to investigate temporal responses in steady states.

Although originally intended to describe the dynamics of interacting and driven-dissipative spin systems, the \ac{twa} can also be applied to calculate thermal and ground states of spin Hamiltonians~\cite{schlegel2026imaginarytimeevolutioninteractingspin} by extension to imaginary times (\ac{itwa}). For general Ising Hamiltonians the \ac{itwa} becomes exact for large inverse temperatures, which has applications to spin-glass physics~\cite{edwards1975theory,parisi1979infinite,binder1986spin} and optimization problems \cite{kirkpatrick1983optimization,du1998handbook,lucas2014ising}. However, while qualitatively correct, it is generally not able to describe the critical regime of quantum phase transitions quantitatively. 

The \ac{twa} for spins is an approximate semiclassical approach which goes beyond the mean-field level. It takes into account leading order quantum effects and can even describe the generation of entanglement. However, it generally does require a truncation of higher-order derivatives in the  \ac{eom} of the Wigner function. In contrast to phase space methods for boson fields, there is no a-priori small parameter justifying such a truncation and the quality of the approximation can only be evaluated a posteriori, e.g. by calculating the coefficients of higher-order derivatives. 
Moreover, stochastic equations can only be derived if the diffusion matrix is positive definite. If this is not the case, diffusion must either be neglected altogether or be approximated. Since dissipative processes generally shift the eigenvalues of the diffusion matrix towards positive values, the \ac{twa} becomes more accurate for increasing dissipative couplings.

In summary we believe that the \ac{twa} for spins in continuous phase space is an extremely versatile and numerically inexpensive method to simulate the dynamics of large interacting, open spin systems.
This also includes problems that can be mapped to coupled spin models, such as the collective emission  of light from quantum emitters.
The \ac{twa} is expected to be rather accurate for systems where quantum correlations are present but are small. Several extensions of the method can be envisioned.
This includes the extension to higher spins.
Furthermore, the degrees of freedom of the TWA -- the gauge freedom of the continuous phase space representation and the choice of correspondence rules -- are not yet fully understood and exploited. Here, a systematic study is needed. Likewise, it would be interesting to explore if short-range quantum correlations can be included exactly, e.g. by cluster-type approaches~\cite{PhysRevB.110.054204} or the combination with \ac{mps} techniques \cite{schollwock}.

\color{black}

\section*{Acknowledgement}
The authors would like to thank Ana Asenjo Garcia, Dennis Breu, Antoine Browaeys, Silvia Fernanda Cardenas Lopez, Igor Ferrier-Barbut, Edgar Guardiola Navarrete, Igor Lesanovsky, Aleksandr Mikheev, Peter Rabl, Johannes Schachenmayer, and Felix Tebbenjohanns for fruitful discussions.
J.H., T.S., C.M. and M.F. acknowledge financial support from the DFG through SFB TR 185, Project No. 277625399. V.N. also acknowledges support by the ERC grant OPEN-2QS (Grant No. 101164443). The authors also thank the Allianz für Hochleistungsrechnen (AHRP) for giving us access to the “Elwetritsch” HPC Cluster through the AHRP project TWEAQING.

\section*{Data availability}
The data that support the findings of this article are not publicly available. The data are available from the authors upon reasonable request.

\appendix

\section{Spherical Harmonics}

\subsection{Definition}

Here we summarize some properties of spherical harmonics and list the definitions used in this paper.
Spherical harmonics \( Y_{l m}(\theta, \phi) \) are a set of special functions defined on the unit sphere \( S^2 \). They serve as the angular portion of the solutions to the Laplace equation in spherical coordinates and form an orthonormal basis for the square-integrable functions on the sphere.

The spherical harmonics are indexed by two integers:
\begin{itemize}
    \item \( l = 0, 1, 2, \ldots \) (the degree or angular momentum quantum number)
    \item \( m = -l, -l+1, \ldots, l \) (the order or magnetic quantum number)
\end{itemize}

They are typically normalized such that
\[
Y_{l m}(\theta, \phi) = \sqrt{\frac{2l+1}{4\pi} \frac{(l-m)!}{(l+m)!}} \, P_l^m(\cos \theta) \, e^{i m \phi},
\]
where \( P_l^m(x) \) are the associated Legendre polynomials.

\subsection{Properties}

\begin{itemize}
    \item \textbf{Orthonormality:}
    \[
    \int_0^\pi \int_0^{2\pi} Y_{l m}^*(\theta, \phi) \, Y_{l' m'}(\theta, \phi) \, \sin \theta \, d\theta \, d\phi = \delta_{l l'} \delta_{m m'}.
    \]

    \item \textbf{Completeness:} Any square-integrable function on the sphere can be expanded uniquely in terms of spherical harmonics.

    \begin{align}
    \sum_{l=0}^\infty \sum_{m=-l}^l \, \Y_l^{m*}(\Omega) \Y_l^{m}(\Omega^\prime) & = 
    \delta(\phi-\phi^\prime)\delta(\cos\theta-\cos\theta^\prime)\nonumber\\
    &= \frac{1}{2\pi}\delta(\Omega-\Omega^\prime)\label{eq:complete}
    \end{align}

    \item \textbf{Conjugation symmetry:}
    \[
    Y_{l m}^*(\theta, \phi) = (-1)^m \, Y_{l, -m}(\theta, \phi).
    \]

    \item \textbf{Transformation under rotation:}

    Under the action of rotation operators parametrized by Euler angles \((\alpha, \beta, \gamma)\), spherical harmonics transform as
    \[
    R(\alpha, \beta, \gamma) \, Y_{l m}(\theta, \phi) = \sum_{m'=-l}^{l} D^{(l)}_{m' m} (\alpha, \beta, \gamma) \, Y_{l m'}(\theta, \phi),
    \]
    where \( D^{(l)}_{m' m} \) are the Wigner D-matrix elements for angular momentum \( l \).
\end{itemize}

\subsection*{Explicit expressions for first few spherical harmonics}

Using the standard Condon-Shortley phase convention, the first few spherical harmonics read:

\paragraph*{\(l=0\):}
\[
Y_{0 0}(\theta, \phi) = \frac{1}{2} \sqrt{\frac{1}{\pi}}.
\]

\paragraph*{\(l=1\):}
\[
Y_{1, -1}(\theta, \phi) = \frac{1}{2} \sqrt{\frac{3}{2\pi}} \sin \theta \, e^{-i \phi},
\]
\[
Y_{1, 0}(\theta, \phi) = \frac{1}{2} \sqrt{\frac{3}{\pi}} \cos \theta,
\]
\[
Y_{1, 1}(\theta, \phi) = -\frac{1}{2} \sqrt{\frac{3}{2\pi}} \sin \theta \, e^{i \phi}.
\]

\paragraph*{\(l=2\):}
\[
Y_{2, -2}(\theta, \phi) = \frac{1}{4} \sqrt{\frac{15}{2\pi}} \sin^2 \theta \, e^{-2 i \phi},
\]
\[
Y_{2, -1}(\theta, \phi) = \frac{1}{2} \sqrt{\frac{15}{2\pi}} \sin \theta \cos \theta \, e^{-i \phi},
\]
\[
Y_{2, 0}(\theta, \phi) = \frac{1}{4} \sqrt{\frac{5}{\pi}} (3 \cos^2 \theta - 1),
\]
\[
Y_{2, 1}(\theta, \phi) = -\frac{1}{2} \sqrt{\frac{15}{2\pi}} \sin \theta \cos \theta \, e^{i \phi},
\]
\[
Y_{2, 2}(\theta, \phi) = \frac{1}{4} \sqrt{\frac{15}{2\pi}} \sin^2 \theta \, e^{2 i \phi}.
\]

\ 

The inverse relations are also useful
\begin{align}
    \cos\theta &= \sqrt{\frac{4\pi}{3}} \Y_1^0(\Omega),\\
    \sin\theta\cos\phi&= \frac{1}{2}\sqrt{\frac{8\pi}{3}}\left(\Y_1^{-1} - \Y_1^1\right) \nonumber\\
    &=\frac{1}{2}\sqrt{\frac{8\pi}{3}}\left(\Y_1^{-1} + \Y_1^{-1*}\right)\\
    & = \sqrt{\frac{8\pi}{3}} \textrm{Re}\left[\Y_1^{-1}(\Omega)\right] = -\sqrt{\frac{8\pi}{3}} \textrm{Re}\left[\Y_1^{1}(\Omega)\right]\nonumber\\
   \sin\theta\sin\phi & = \frac{i}{2}\sqrt{\frac{8\pi}{3}}\left(\Y_1^{-1} + \Y_1^1
\right) \nonumber\\
&=\frac{i}{2}\sqrt{\frac{8\pi}{3}}\left(\Y_1^{-1} - \Y_1^{-1*}\right)\\
& = -\sqrt{\frac{8\pi}{3}} \textrm{Im}\left[\Y_1^{-1}(\Omega)\right] = -\sqrt{\frac{8\pi}{3}} \textrm{Im}\left[\Y_1^{1}(\Omega)\right]\nonumber
\end{align}

\section{Relation between Pauli matrices and phase point operators }

For the mapping of the density matrix equations to a PDE of $W(\Om,t)$ the following relations between Pauli matrices and (derivatives of the) phase point operator are useful. Starting from

\begin{equation*}
    \hat{\Delta} = \frac{\sqrt{3}}{2}\Bigl(\sin\theta\cos\phi\, \sigma_x +\sin\theta\sin\phi \,\sigma_y + \cos\theta \,\sigma_z\Bigr) +\frac{1}{2}\mathbf{1}
\end{equation*}
one finds
\begin{eqnarray}
    \partial_\theta \hat{\Delta} &=& \frac{\sqrt{3}}{2} \Bigl(\cos\theta\cos\phi \,\sigma_x +\cos\theta\sin\phi\, \sigma_y - \sin\theta \,\sigma_z\Bigr),\nonumber\\
    \partial_\phi\hat{\Delta} &=& \frac{\sqrt{3}}{2} \Bigl(-\sin\theta\sin\phi \,\sigma_x +\sin\theta\cos\phi\, \sigma_y \Bigr),\\
    \partial^2_\phi\hat{\Delta} &=& \frac{\sqrt{3}}{2} \Bigl(-\sin\theta\cos\phi \,\sigma_x -\sin\theta\sin\phi\, \sigma_y \Bigr).\nonumber
\end{eqnarray}
which can easily be inverted for the Pauli matrices
\begin{eqnarray}
    \mathbf{1} &=& 2\hat{\Delta} + 2\cot\theta \, \partial_\theta \hat{\Delta} + 2\csc^2\theta \, \partial_\phi^2\hat{\Delta},\nonumber\\
    \sigma_x &=& -\frac{2 \csc\theta }{\sqrt{3}} \left(\sin\phi \, \partial_\phi \hat{\Delta}+ \cos\phi \, \partial^2_\phi \hat{\Delta}\right),\nonumber\\
     \sigma_y &=& \frac{2\csc\theta }{\sqrt{3}} \left( \cos\phi \, \partial_\phi \hat{\Delta}-\sin\phi \, \partial^2_\phi \hat{\Delta}\right),\\
      \sigma_z &=& \frac{2}{\sqrt{3}} \left(
     -\csc\theta  \,  \partial_\theta \hat{\Delta} - \csc\theta \cot\theta  \, \partial^2_\phi  \hat{\Delta} \right).\nonumber
\end{eqnarray}
or in more compact form
\begin{eqnarray}
    \sigma_\pm &=& \frac{1}{2}(\sigma_x\pm i \sigma_y) \nonumber\\
    &=&\frac{1}{\sqrt{3}} \csc\theta e^{\pm i\phi} \Bigl(\pm i \partial_\phi \hat{\Delta}- \partial^2_\phi\hat{\Delta}\Bigr)\\
     \sigma_z &=& - \frac{2}{\sqrt{3}} \csc\theta \left(
     \partial_\theta \hat{\Delta} + \cot\theta  \, \partial^2_\phi  \hat{\Delta} \right).\nonumber
\end{eqnarray}

\

\color{black}

\vfill

\bibliography{dissipativeDTWA.bib}

\end{document}